\documentclass[twocolumn,amsmath,amssymb,amsfonts,floatfix,superscriptaddress]{revtex4}

\usepackage{graphicx}
\usepackage{lscape}
\usepackage{float}
\usepackage{psfrag}
\usepackage{algpseudocode}
\usepackage{appendix}
\usepackage{caption}
\usepackage{subfig}

\begin{document}

\title{Propagative Block Diagonalisation Diabatisation of DFT/MRCI
  Electronic States}

\author{Simon P. Neville}
\email{simon.neville@nrc-cnrc.gc.ca}
\altaffiliation{Current Address: National Research Council of Canada,
  100 Sussex Drive, Ottawa, Ontario K1A 0R6, Canada}
\affiliation{Department of Chemistry and Biomolecular Sciences,
  University of Ottawa, 10 Marie Curie, Ottawa, Ontario, K1N 6N5,
  Canada}

\author{Issaka Seidu}
\affiliation{Department of Chemistry and Biomolecular Sciences,
  University of Ottawa, 10 Marie Curie, Ottawa, Ontario, K1N 6N5,
  Canada}

\author{Michael S. Schuurman}
\affiliation{Department of Chemistry and Biomolecular Sciences,
  University of Ottawa, 10 Marie Curie, Ottawa, Ontario, K1N 6N5,
  Canada}
\affiliation{National Research Council of Canada, 100 Sussex Drive,
  Ottawa, Ontario K1A 0R6, Canada}

\begin{abstract}
  We present a framework for the calculation of diabatic states using
  the combined density functional theory and multireference
  configuration interaction (DFT/MRCI) method. Due to restrictions
  present in the current formulation of the DFT/MRCI method (a lack of
  analytical derivative couplings and the inability to use
  non-canonical Kohn-Sham orbitals), most common diabatisation
  strategies are not applicable. We demonstrate, however, that
  diabatic wavefunctions and potentials can be calculated at the
  DFT/MRCI level of theory using a propagative variant of the block
  diagonalisation diabatisation method (P-BDD). The proposed procedure
  is validated {\it via} the calculation of diabatic potentials for
  LiH and the simulation of the vibronic spectrum of pyrazine. In both
  cases, the combination of the DFT/MRCI and P-BDD methods is found to
  correctly recover the non-adiabatic coupling effects of the problem.
\end{abstract}

\maketitle

\section{Introduction}
Since its introduction by Grimme and Waletzke\cite{grimme_dft-mrci},
the combined density functional theory and multireference
configuration interaction (DFT/MRCI) method has proved to be a
uniquely powerful general purpose semi-empirical method for the
calculation of the excited states of large molecules. Advantages of
the DFT/MRCI method include its accuracy, computational efficiency,
black box nature, and its ability to correctly describe a wide range
of classes of excited states, including those of valence, Rydberg,
charge transfer, and doubly-excited character. Although originally
conceived as a method for the description of singlet and triplet
states in organic molecules, recent developments by Marian and
co-workers have resulted in a spin multiplicity-independent redesign
of the DFT/MRCI
Hamiltonian\cite{lyskov_dftmrci_redesign,heil_dftmrci_2017}, and a
re-parameterisation for an improved description of transition metal
complexes\cite{heil_dftmrci_transition_metals}.

The DFT/MRCI method has seen extensive use in the characterization of
excited state surfaces including the calculation of excitation
energies, spin-orbit couplings and photochemical reaction
pathways\cite{marian_furan,marian_thiophene,marian_adenine,marian_porphyrin_intersystem_crossing,pyrrole_dimer_ours,marian_dftmrci_review}. One
area where DFT/MRCI has seen less use, however, is in conjunction with
time-dependent quantum dynamics calculations. The starting point for
such studies is the extraction of potential energy surfaces and
non-adiabatic couplings in a suitable form. This may either be in the
adiabatic representation, where the couplings between the electronic
states are described by derivative coupling matrix elements, or in a
diabatic representation, in which the interstate couplings are
described by off-diagonal elements of the potential matrix. Due to
singularities present in the derivative coupling terms in the
adiabatic representation, it is often preferable to turn to a diabatic
representation, where they are eliminated. Furthermore, analytic
derivative couplings are not available at the DFT/MRCI level of
theory. Thus, one is naturally prompted to explore methods for the
calculation of diabatic potential matrices using DFT/MRCI.

For polyatomic molecules, the generation of strictly diabatic states
(which exactly remove the derivative couplings) is not
possible\cite{mead_diabatic}. Instead, one seeks a set of so-called
quasi-diabatic states that minimise the derivative coupling terms, or
at least remove the singular components of them. There is no unique
solution to this problem, and, accordingly, a large number of
diabatisation schemes have been developed over the years. For an
overview of some of representative examples, we direct the reader to
References
\citenum{atchity_configurational_uniformity,truhlar_four-fold_diabatisation,truhlar_four-fold_diabatisation_mcqdpt2,koppel_regularised_diabatic_states_fd,koppel_regularised_diabatic_states_general,pacher_bdd_1989,pacher_quasidiabatic_states_adv_chem_phys,robertson_diabatisation_ras,richings_diabatisation_ddvmcg,truhlar_dq_diabatisation,yarkony_diabatic_potential_fitting,yarkony_neural_networks_diabatic_potentials}. However,
many of the existing methods are not compatible with the current
formulation of DFT/MRCI. The reasons for this are two-fold. Firstly,
the inability to compute analytic derivative couplings means that all
of the methods employing
them\cite{koppel_regularised_diabatic_states_fd,koppel_regularised_diabatic_states_general,richings_diabatisation_ddvmcg,yarkony_diabatic_potential_fitting,yarkony_neural_networks_diabatic_potentials}
cannot be leveraged. Secondly, DFT/MRCI requires the use of the
canonical Kohn-Sham (KS) orbitals as the single-particle basis,
rendering it incompatible with any method that uses bases of diabatic
molecular orbitals
(MOs)\cite{atchity_configurational_uniformity,truhlar_four-fold_diabatisation,truhlar_four-fold_diabatisation_mcqdpt2,robertson_diabatisation_ras}. In
fact, beyond property-based diabatisation
methods\cite{truhlar_dq_diabatisation} and strategies based on
diabatisation by {\it ansatz} (involving the fitting of pre-specified
functional forms to adiabatic energies), we are not aware of any
currently implemented diabatisation schemes that are suitable for use
with DFT/MRCI.

To proceed to a solution to this problem, we re-visit the block
diagonalisation diabatisation (BDD) method of Pacher, Cederbaum and
K\"{o}ppel\cite{pacher_bdd_1989,pacher_quasidiabatic,pacher_quasidiabatic_states_adv_chem_phys}. The
BDD method requires only wavefunction information, making it
compatible with DFT/MRCI. Previous {\it ab initio} implementations of
this method have, however, been based around the construction of
configuration state functions (CSFs) in a diabatic MO
basis\cite{domcke_diabatic_casscf,domcke_diabatic_casscf_ozone,eisfeld_block_diagonalisation_diabatisation,eisfeld_hybrid_block_diagonalisation_diabatisation,stanton_eom_block_diagonalisation},
which is not possible with DFT/MRCI. Instead, we consider an idea
explored by Pacher, Cederbaum and K\"{o}ppel many years
ago\cite{pacher_quasidiabatic_states_adv_chem_phys}, but never before
implemented in an {\it ab initio} study. Specifically, we implement a
propagative BDD (P-BDD) algorithm that requires only the overlaps of
adiabatic electronic wavefunctions at neighboring nuclear
geometries. In this implementation, the BDD method is directly
compatible with DFT/MRCI.

The rest of this paper is organised as follows. In Section
\ref{sec:theory}, we give a discussion of the P-BDD algorithm and the
details of an efficient wavefunction overlap scheme that renders it
computationally tractable. In Section \ref{sec:results}, we assess the
quality of the diabatic potentials yielded by the combination of
DFT/MRCI and P-BDD methods. Firstly, the diabatic potentials of the
first two $^{1}\Sigma^{+}$ states of LiH were calculated and compared
to the results of canonical MRCI computations. Secondly, a model
vibronic coupling Hamiltonian for pyrazine was parameterised using the
results of DFT/MRCI P-BDD calculations and used in the simulation of
its absorption spectrum. In the case of LiH, the DFT/MRCI P-BDD
diabatic potentials were found to yield derivative couplings in good
agreement with the canonical MRCI results. For pyrazine, the simulated
absorption spectrum was found to be in excellent agreement with its
experimental counterpart. Through these results, we conclude that the
combination of the DFT/MRCI and P-BDD methods offers a reliable,
near-black box route to the calculation of diabatic potentials.

\section{Theoretical framework}\label{sec:theory}

\subsection{Propagative block diagonalisation diabatisation}
We begin with the introduction of some notation. In the following,
$\boldsymbol{r}$ and $\boldsymbol{R}$ are used to denote the vectors
of electronic and nuclear coordinates, respectively. Let $\hat{H}$
denote the total molecular Hamiltonian, partitioned as

\begin{equation}\label{eq:molecular_ham}
  \hat{H}(\boldsymbol{r}, \boldsymbol{R}) =
  \hat{T}_{n}(\boldsymbol{R}) +
  \hat{H}_{el}(\boldsymbol{r},\boldsymbol{R}),
\end{equation}

\noindent
where $\hat{T}_{n}(\boldsymbol{R})$ is the nuclear kinetic energy
operator and $\hat{H}_{el}(\boldsymbol{r}, \boldsymbol{R})$ the
electronic Hamiltonian, that is, the sum of the electronic and nuclear
kinetic energy operator and all Coulombic potential terms. The set of
eigenfunctions of $\hat{H}_{el}(\boldsymbol{r}, \boldsymbol{R})$ (the
set of adiabatic electronic states) is denoted by $\{
\psi_{I}(\boldsymbol{r}; \boldsymbol{R}) \}$. The corresponding
eigenvalues (the adiabatic potentials) are denoted by
$V_{I}(\boldsymbol{R})$. Our goal is to compute a set $\{
\phi_{I}(\boldsymbol{r}; \boldsymbol{R}) \}$ of quasi-diabatic states,
related to the adiabatic states through the adiabatic-to-diabatic
transformation (ADT) matrix $\boldsymbol{U}(\boldsymbol{R})$:

\begin{equation}
  \phi_{I}(\boldsymbol{r}; \boldsymbol{R}) = \sum_{J}
  U_{JI}(\boldsymbol{R}) \psi_{J}(\boldsymbol{r}; \boldsymbol{R}).
\end{equation}

\noindent
Finally, let $\{ \Phi_{I}(\boldsymbol{r}; \boldsymbol{R}) \}$ be a set
of `initial', or `reference', states in terms of which the Hamiltonian
is to be initially represented. More will subsequently be said about
the possible choices for the initial states, but for now they are left
unspecified beyond being taken to form a complete orthonormal basis in
which the molecular wavefunction can be represented.

A key idea underlying the BDD method is that, within the framework of
the group Born-Oppenheimer approximation, it is usually possible to
identify a finite subset $P$ of states of interest that are well
separated from, and only weakly coupled to, their orthogonal
complement of states, termed the $Q$ space states. We may then seek an
ADT matrix $\boldsymbol{U}(\boldsymbol{R})$ that operates only in the
$P$ space of states, i.e., with the structure

\begin{equation}
  \boldsymbol{U}(\boldsymbol{R}) =
  \left(
  \begin{array}{cc}
    \boldsymbol{U}_{P}(\boldsymbol{R}) & \boldsymbol{0} \\
    \boldsymbol{0} & \boldsymbol{1}
  \end{array}
  \right),
\end{equation}

\noindent
where $\boldsymbol{U}_{P}(\boldsymbol{R})$ denotes the $P$ space part
of $\boldsymbol{U}(\boldsymbol{R})$. It is obvious that any such $P$
space transformation will result in a representation of the electronic
Hamiltonian that is block diagonal in the $P$ and $Q$ spaces. Knowing
that a quasi-diabatisation within the $P$ space will yield a block
diagonal electronic Hamiltonian, a natural question to ask is whether
we can determine a block diagonalisation transformation of the
electronic Hamiltonian matrix that leads to quasi-diabatic states?
This is exactly the issue addressed by Pacher, Cederbaum and K\"{o}ppel
in their BDD
scheme\cite{pacher_bdd_1989,pacher_quasidiabatic,pacher_quasidiabatic_states_adv_chem_phys}.

The electronic Hamiltonian is first represented using the basis $\{
\Phi_{I}(\boldsymbol{r};\boldsymbol{R}) \}$ of initial states:

\begin{equation}
  H_{IJ}(\boldsymbol{R}) = \left\langle
  \Phi_{I}(\boldsymbol{r};\boldsymbol{R}) \middle| \hat{H}_{el} \middle|
  \Phi_{J}(\boldsymbol{r};\boldsymbol{R}) \right\rangle.
\end{equation}

\noindent
A unitary transformation $\boldsymbol{B}(\boldsymbol{R})$ is sought
that brings the initial state electronic Hamiltonian matrix
$\boldsymbol{H}(\boldsymbol{R})$ into the desired block diagonal form,
denoted by $\boldsymbol{\mathcal{H}}(\boldsymbol{R})$:

\begin{equation}
  \begin{aligned}
    \boldsymbol{\mathcal{H}}(\boldsymbol{R}) &=
    \boldsymbol{B}^{\dagger}(\boldsymbol{R})
    \boldsymbol{H}(\boldsymbol{R}) \boldsymbol{B}(\boldsymbol{R}) \\
    &= \left(
    \begin{array}{cc}
      \boldsymbol{\mathcal{H}}_{P} & \boldsymbol{0} \\
      \boldsymbol{0} & \boldsymbol{\mathcal{H}}_{Q}
    \end{array}
    \right).
  \end{aligned}
\end{equation}

\noindent
There are an infinite number of transformations
$\boldsymbol{B}(\boldsymbol{R})$ that will result in the desired block
diagonal structure, and a constraint must be introduced. In
particular, in the BDD scheme, the constraint used takes the form of a
`least action principle' of the form

\begin{equation}\label{eq:bdd_constraint}
  || \boldsymbol{B}(\boldsymbol{R}) - \boldsymbol{1} || = min,
\end{equation}

\noindent
where $||\cdot||$ denotes the Frobenius norm. That is, the
transformation $\boldsymbol{B}(\boldsymbol{R})$ is constrained to
block diagonalise the initial state electronic Hamiltonian matrix
$\boldsymbol{H}(\boldsymbol{R})$, but beyond that do as little as
possible. With this one constraint in place, the transformation
$\boldsymbol{B}(\boldsymbol{R})$ may be uniquely determined to take
the following
form\cite{cederbaum_block_diagonalisation,pacher_bdd_1989}:

\begin{equation}
  \boldsymbol{B}(\boldsymbol{R}) = \boldsymbol{Z}(\boldsymbol{R})
  \left( \boldsymbol{Z}^{\dagger}(\boldsymbol{R})
  \boldsymbol{Z}(\boldsymbol{R}) \right)^{-\frac{1}{2}},
\end{equation}

\noindent
where

\begin{equation}
  \boldsymbol{Z}(\boldsymbol{R}) = \boldsymbol{S}(\boldsymbol{R})
  \boldsymbol{S}_{BD}^{-1}(\boldsymbol{R}).
\end{equation}

\noindent
Here, $\boldsymbol{S}(\boldsymbol{R})$ denotes the matrix over
overlaps between the initial and adiabatic electronic states,

\begin{equation}
  S_{IJ}(\boldsymbol{R}) = \left\langle
  \Phi_{I}(\boldsymbol{r};\boldsymbol{R}) \middle|
  \psi_{J}(\boldsymbol{r};\boldsymbol{R}) \right\rangle,
\end{equation}

\noindent
and $\boldsymbol{S}_{BD}(\boldsymbol{R})$ is its block diagonal part:

\begin{equation}
  \boldsymbol{S}_{BD}(\boldsymbol{R}) = \left(
  \begin{array}{cc}
    \boldsymbol{S}_{P}(\boldsymbol{R}) & \boldsymbol{0} \\
    \boldsymbol{0} & \boldsymbol{S}_{Q}(\boldsymbol{R})
  \end{array}
  \right).
\end{equation}

Finally, by making use of the relation

\begin{equation}
  \boldsymbol{B}(\boldsymbol{R}) = \boldsymbol{S}(\boldsymbol{R})
  \boldsymbol{U}(\boldsymbol{R}),
\end{equation}

\noindent
the ADT matrix $\boldsymbol{U}(\boldsymbol{R})$ may be written as

\begin{equation}
  \boldsymbol{U}(\boldsymbol{R}) =
  \boldsymbol{S}_{BD}^{-1}(\boldsymbol{R})
  (\boldsymbol{S}_{BD}(\boldsymbol{R})
  \boldsymbol{S}_{BD}^{\dagger}(\boldsymbol{R}))^{\frac{1}{2}}.
\end{equation}

\noindent
Importantly, the $P$ space part of $\boldsymbol{U}(\boldsymbol{R})$
that we require can be computed using only the $P$ space part of
$\boldsymbol{S}_{BD}(\boldsymbol{R})$. That is, using only the overlaps
of the initial and adiabatic electronic states spanning the $P$
space:

\begin{equation}\label{eq:bdd_adt}
  \boldsymbol{U}_{P}(\boldsymbol{R}) =
  \boldsymbol{S}_{P}^{-1}(\boldsymbol{R})
  (\boldsymbol{S}_{P}(\boldsymbol{R})
  \boldsymbol{S}_{P}^{\dagger}(\boldsymbol{R}))^{\frac{1}{2}}.
\end{equation}

\noindent
All we then require is a prescription for the choice of the initial
states $\Phi_{I}(\boldsymbol{r};\boldsymbol{R})$.

In order for the transformation $\boldsymbol{B}(\boldsymbol{R})$ to
rotate the initial basis into a good set of quasi-diabatic states $\{
\phi_{I}(\boldsymbol{r;R}) \}$, it is required that the initial states
$\Phi_{I}(\boldsymbol{r;R})$ themselves already behave somewhat
quasi-diabatically, or at least vary smoothly with the nuclear
geometry, as can be deduced from the least action principle, Equation
\ref{eq:bdd_constraint}. Soon after the development of the BDD method,
it was recognised that a convenient choice of such a set of initial
states is the set of CSFs used to expand the adiabatic states
represented in terms of diabatic
MOs\cite{domcke_diabatic_casscf,domcke_diabatic_casscf_ozone}. That
is, MOs obtained from the canonical set by a unitary transformation to
maximise their overlap with the canonical MOs at some reference
geometry $\boldsymbol{R}_{0}$. In practical {\it ab initio}
calculations, this turns out to be an excellent choice as
$\boldsymbol{S}_{P}(\boldsymbol{R})$ then reduces to the matrix of
expansion coefficients for the electronic states in the basis of the
CSFs formed from the diabatic MOs. Indeed, this approach has proved to
be very successful, and implementations have been reported at the
complete active space self consistent field
(CASSCF)\cite{domcke_diabatic_casscf,domcke_diabatic_casscf_ozone,eisfeld_block_diagonalisation_diabatisation},
multi-reference configuration interaction
(MRCI)\cite{eisfeld_hybrid_block_diagonalisation_diabatisation} and
equation-of-motion coupled cluster
(EOM-CC)\cite{stanton_eom_block_diagonalisation} levels of theory.

However, the use of diabatic CSFs as the initial states is not
compatible with DFT/MRCI, which requires the use of the canonical KS
orbitals. Instead, we re-visit a suggestion first put forwards by
Pacher, Cederbaum and K\"{o}ppel in Reference
\citenum{pacher_quasidiabatic_states_adv_chem_phys}, but, as far as we
are aware, never implemented beyond model studies. The idea is to
generate an initial basis $\{ \Phi_{I}(\boldsymbol{r};\boldsymbol{R})
\}$ in a propagative manner. Firstly, a reference geometry
$\boldsymbol{R}_{0}$ is defined, at which the adiabatic and diabatic
representations are taken to be equal, fixing the global gauge of the
adiabatic-to-diabatic transformation. Next, a string of displaced
geometries $\boldsymbol{R}_{n} = \boldsymbol{R}_{0} + n \Delta
\boldsymbol{R}$ is chosen, at which the ADT matrix is to be
calculated. At the first geometry, $\boldsymbol{R}_{1}$, the adiabatic
states $\{ \psi_{I}(\boldsymbol{r};\boldsymbol{R}_{0}) \}$ of the
reference geometry are used as the initial states. That is, $\{
\Phi_{I}(\boldsymbol{r};\boldsymbol{R}_{1}) \} = \{
\psi_{I}(\boldsymbol{r};\boldsymbol{R}_{0}) \}$. This is equivalent to
using a crude adiabatic basis as the initial states at
$\boldsymbol{R}_{1}$. The overlaps between the sets of states $\{
\psi_{I}(\boldsymbol{r};\boldsymbol{R}_{0}) \}$ and $\{
\psi_{I}(\boldsymbol{r};\boldsymbol{R}_{1}) \}$ are calculated, and
used in the construction of the $P$ space ADT matrix
$\boldsymbol{U}_{P}(\boldsymbol{R}_{1})$, as given in Equation
\ref{eq:bdd_adt}. At the next geometry, $\boldsymbol{R}_{2}$, the
quasi-diabatic states $\{ \phi_{I}(\boldsymbol{r};\boldsymbol{R}_{1})
\}$ from $\boldsymbol{R}_{1}$, calculated using
$\boldsymbol{U}_{P}(\boldsymbol{R}_{1})$, are used as the initial
states, yielding the ADT matrix
$\boldsymbol{U}_{P}(\boldsymbol{R}_{2})$. This process is repeated,
using the quasi-diabatic states from the geometry $\boldsymbol{R}_{n}$
as the initial states for the geometry $\boldsymbol{R}_{n+1}$.

The above described propagative BDD, or P-BDD, scheme leads to an
optimal set of quasi-diabatic states $\{
\phi_{I}(\boldsymbol{r};\boldsymbol{R}) \}$ in the limit of
infinitesimal displacements $\Delta \boldsymbol{R}$. Here, the term
optimal is used in the sense of minimising the integral of the squared
norm of the $P$ space derivative coupling matrix over the path
connecting the nuclear coordinates
$\boldsymbol{R}_{0},\boldsymbol{R}_{1},\dots,\boldsymbol{R}_{n}$ (see
Reference \citenum{pacher_quasidiabatic_states_adv_chem_phys} for
details). However, the P-BDD method has not yet seen use beyond the
solution of model diabatisation
problems\cite{pacher_quasidiabatic_states_adv_chem_phys}. This seems
to be rooted in the potential difficulties associated with the
efficient calculation of overlaps of electronic wavefunctions computed
at different nuclear geometries, i.e., using non-orthogonal sets of
MOs. In Section \ref{sec:wf_overlaps}, we describe our approach to
dealing with this problem.

\subsection{Efficient calculation of wavefunction
  overlaps}\label{sec:wf_overlaps}
In order to calculate the ADT matrix of the P-BDD scheme, we have to
compute the overlap matrix elements

\begin{equation}
  \begin{aligned}
    S_{IJ}(\boldsymbol{R}_{n+1}) &= \left\langle
    \phi_{I}(\boldsymbol{r};\boldsymbol{R}_{n}) \middle|
    \psi_{J}(\boldsymbol{r};\boldsymbol{R}_{n+1}) \right\rangle \\ &=
    \sum_{K} U_{IK}^{\dagger}(\boldsymbol{R}_{n}) \left\langle
    \psi_{K}(\boldsymbol{r};\boldsymbol{R}_{n}) \middle|
    \psi_{J}(\boldsymbol{r};\boldsymbol{R}_{n+1}) \right\rangle.
  \end{aligned}
\end{equation}

\noindent
That is, only the overlaps between adiabatic electronic states at
neighboring geometries are needed as input. This is compatible with
the DFT/MRCI method. However, the success of this scheme relies on the
ability to rapidly compute overlaps between wavefunctions expressed in
terms of non-orthogonal orbitals. Careful attention has to be paid
here as, if naively implemented, this can be disastrously slow for
large wavefunction expansions. Our strategy here is based on a
modification of the algorithm originally presented by Plasser {\it et
  al.}\cite{plasser_wf_overlaps}.

We consider the calculation of the set $\{ \langle \psi_{I} |
\psi_{J}' \rangle \}$ of all possible overlaps between two sets of
electronic states $\{ \psi_{I} \}$ and $\{ \psi_{I}' \}$. Here, we
drop the explicit labeling of the coordinate dependence of the
electronic states to avoid unwieldy expressions. In the context of a
P-BDD calculation, the two sets of states would be the adiabatic
electronic states at two neighboring geometries. The states in the
sets $\{ \psi_{I} \}$ and $\{ \psi_{I}' \}$ are in turn expanded in
terms of sets of Slater determinants $\{ \Theta_{k} \}$ and $\{
\Theta_{k}' \}$:

\begin{equation}
  \psi_{I} = \sum_{k=1}^{N_{I}} C_{kI} \Theta_{k},
\end{equation}

\begin{equation}
  \psi_{I}' = \sum_{k=1}^{N_{I}'} C_{kI}' \Theta_{k}'.
\end{equation}

\noindent
Finally, the Slater determinants are expressed in terms of sets of
spin orbitals $\{ \varphi_{n},\bar{\varphi}_{n} \}$ and $\{
\varphi_{n}',\bar{\varphi}_{n}' \}$:

\begin{equation}
  \Theta_{k} = \left| \varphi_{i_{1}^{k}} \cdots
  \varphi_{i_{n_{\alpha}}^{k}} \bar{\varphi}_{i_{1}^{k}} \cdots
  \bar{\varphi}_{i_{n_{\beta}}^{k}} \right|,
\end{equation}

\begin{equation}
  \Theta_{k}' = \left| \varphi_{i_{1}^{k}}' \cdots
  \varphi_{i_{n_{\alpha}}^{k}}' \bar{\varphi}_{i_{1}^{k}}' \cdots
  \bar{\varphi}_{i_{n_{\beta}}^{k}}' \right|.
\end{equation}

\noindent
Here, $\varphi$ and $\bar{\varphi}$ are used to denote $\alpha$ and
$\beta$ spin orbitals, respectively, and $n_{\alpha}$ and $n_{\beta}$
the number of $\alpha$ and $\beta$ electrons. The multi-index
$i_{m}^{k}$ is used to denote the index of the $\alpha$ ($\beta$) spin
orbital that is in position $m$ ($n_{\alpha}+m$) in the $k$th Slater
determinant $\Theta_{k}$.

The overlap between two states $\psi_{I}$ and $\psi_{J}'$ is then
given by

\begin{equation}\label{eq:wf_overlap}
  \left\langle \psi_{I} \middle| \psi_{J}' \right\rangle =
  \sum_{k=1}^{N_{I}} \sum_{l=1}^{N_{J}'} C_{kI} C_{lJ}' \left\langle
  \Theta_{k} \middle| \Theta_{l}' \right\rangle.
\end{equation}

\noindent
The bottleneck in the evaluation of Equation \ref{eq:wf_overlap} is
the calculation of the overlaps between the Slater determinants in
the sets $\{ \Theta_{k} \}$ and $\{ \Theta_{k}' \}$, which are given
by the determinant of the matrix of overlaps between the spin orbitals
occupied in the two Slater determinants:

\begin{widetext}

  \begin{equation}\label{eq:det_overlap}
    \left\langle \Theta_{k} \middle| \Theta_{l}' \right\rangle = \left|
    \begin{array}{cccccc}
      \langle \varphi_{i_{1}^{k}} | \varphi_{i_{1}^{l}}' \rangle & \cdots &
      \langle \varphi_{i_{n_{\alpha}}^{k}} | \varphi_{i_{n_{\alpha}}^{l}}'
      \rangle &  &  &  \\
      \vdots & \ddots & \vdots &  & \boldsymbol{0} &  \\
      \langle \varphi_{i_{n_{\alpha}}^{k}} | \varphi_{i_{1}^{l}}' \rangle &
      \cdots & \langle \varphi_{i_{n_{\alpha}}^{k}} |
      \varphi_{i_{n_{\alpha}}^{l}}' \rangle & & & \\
      & & & \langle \bar{\varphi}_{i_{1}^{k}} | \bar{\varphi}_{i_{1}^{l}}'
      \rangle & \cdots & \langle \bar{\varphi}_{i_{1}^{k}} |
      \bar{\varphi}_{i_{n_{\beta}}^{l}}' \rangle \\
      & \boldsymbol{0} & & \vdots & \ddots & \vdots \\
      & & & \langle \bar{\varphi}_{i_{n_{\beta}}^{k}} | \bar{\varphi}_{i_{1}^{l}}'
      \rangle & \cdots & \langle \bar{\varphi}_{i_{n_{\beta}}^{k}} |
      \bar{\varphi}_{i_{n_{\beta}}^{l}}' \rangle \\
    \end{array}
    \right|
    =\left|
    \begin{array}{cc}
      \boldsymbol{s}_{kl} & \boldsymbol{0} \\
      \boldsymbol{0} & \bar{\boldsymbol{s}}_{kl} \\
    \end{array}
    \right|  = \det(\boldsymbol{s}_{kl}) \times \det(
    \bar{\boldsymbol{s}}_{kl})
  \end{equation}

\end{widetext}

If directly implemented, with the calculation of the factors
$\det(\boldsymbol{s}_{kl})$ and $\det(\bar{\boldsymbol{s}}_{kl})$
being performed on-the-fly for every pair of Slater determinants, the
computational effort for the calculation of the wavefunction overlap
$\langle \psi_{I} | \psi_{J} \rangle$ scales as $N_{I}
N_{J}'n_{el}^{3}$, where $N_{I}$ and $N_{J}'$ are the number of Slater
determinants in the expansion of $\psi_{I}$ and $\psi_{J}'$,
respectively, and $n_{el}$ is the number of electrons. For DFT/MRCI
wavefunctions, the dimension of the Slater determinant basis is
usually $\mathcal{O}(10^{4}-10^{6})$, resulting in potentially ruinous
computational costs.

A way to alleviate this bottleneck was recently put forward by Plasser
{\it et al.}\cite{plasser_wf_overlaps}. As the authors noted, the
factors $\det(\boldsymbol{s}_{kl})$ and
$\det(\bar{\boldsymbol{s}}_{kl})$ are not unique to the pair of Slater
determinants $\Theta_{k}$ and $\Theta_{l}'$, but also occur for other
Slater determinants with the same $\alpha$ and $\beta$ spin orbital
occupations. By precomputing and storing all unique factors for the
wavefunction pair $\psi_{I}$ and $\psi_{J}'$, speedups of many orders
of magnitude can be attained. Our approach is based on the algorithm
reported in Reference \citenum{plasser_wf_overlaps}, but with two key
modifications.

Firstly, we note that common $\alpha$ and $\beta$ spin
orbital occupations occur not just between pairs of wavefunctions
$\psi_{I}$ and $\psi_{J}'$, but between sets of wavefunctions $\{
\psi_{I} \}$ and $\{ \psi_{I}' \}$. Thus, we identify, precompute and
store all unique factors $\det(\boldsymbol{s}_{kl})$ and $\det(
\bar{\boldsymbol{s}}_{kl})$ based on the common $\alpha$ and $\beta$
spin orbital occupations across all states in the sets $\{ \psi_{I}
\}$ and $\{ \psi_{I}' \}$. In the P-BDD procedure,
$n_{state}^{2}$ wavefunction overlaps have to be computed, where
$n_{state}$ is the number of electronic states being considered. For
even moderate numbers of states, precomputing the unique factors
across all pairs of states, instead of on a pair-by-pair basis, can
result in significant computational savings.

Secondly, we recognise that many of the unique factors $\det(
\boldsymbol{s}_{kl})$ and $\det(\bar{\boldsymbol{s}}_{kl})$ will be
very small in magnitude and can, in fact, be neglected without
detriment. We thus introduce a screening step in the generation of the
unique factors $\det(\boldsymbol{s}_{kl})$ and $\det(
\bar{\boldsymbol{s}}_{kl})$. This requires a fast, robust estimate of
the magnitude of a determinant. For this we make use of Hadamard's
inequality, which, for a matrix $\boldsymbol{X} \in \mathbb{R}^{m
  \times m}$, reads

\begin{equation}\label{eq:hadamard}
  \left| \det(\boldsymbol{X}) \right|^{2} \le \prod_{i=1}^{m}
  ||\boldsymbol{x}_{i}||^{2},
\end{equation}

\noindent
where $\boldsymbol{x}_{i}$ are the columns of $\boldsymbol{X}$. The
calculation of the bound Equation \ref{eq:hadamard} scales as
$\mathcal{O}(m^{2})$, whereas the calculation of the determinant
scales as $\mathcal{O}(m^{3})$. For the factors
$\det(\boldsymbol{s}_{kl})$ and $\det( \bar{\boldsymbol{s}}_{kl})$,
the matrix dimension, $m$, is $n_{\alpha}$ and $n_{\beta}$,
respectively. Thus, if many of the factors are small enough to be
neglected, then screening these using Hadamard's inequality will lead
to a considerable speedup. We refer to this screening step as Hadamard
screening.

Lastly, we also consider the use of truncation as a way to speed up
the computation of the wavefunction overlaps. In particular, a
norm-based truncation is used in which the replacement

\begin{equation}
  \psi_{I} = \sum_{k=1}^{N_{I}} C_{kI} \Theta_{k} \rightarrow
  \tilde{\psi}_{I} = \sum_{k \in \mathcal{S}_{I}} \tilde{C}_{kI}
  \Theta_{k},
\end{equation}

\begin{equation}
  \tilde{C}_{kI} = \sqrt{\sum_{k \in \mathcal{S}_{I}} C_{kI}^{2}},
\end{equation}

\noindent
is made. Here, $\mathcal{S}_{I}$ is the smallest subset of Slater
determinant indices that yields a truncated wavefunction with a norm
above a given threshold $\delta_{t}$ for the $I$th state.

\section{Methodology}
\subsection{Calculation of the vibronic absorption spectrum of
  pyrazine}
As part of our analysis of the quality of the diabatic potentials
yielded by the DFT/MRCI P-BDD method, we explored the construction of
model potentials for use in the simulation of vibronic spectra. The
test case chosen was the absorption spectrum of pyrazine. This is an
often-used benchmark system for the study of vibronic coupling effects
in absorption spectra, with a proper description of the strong
coupling between its first two excited states being necessary for the
correct reproduction of the absorption spectrum.

Let $\sigma_{I}(E)$ denote the absorption spectrum corresponding to
vertical excitation from the ground state to the excited diabatic
electronic state
$\phi_{I}(\boldsymbol{r};\boldsymbol{R})$. $\sigma_{I}(E)$ may be
calculated within a time-dependent framework from the Fourier
transform of the autocorrelation function $a_{I}(t)$ corresponding to
an initial wavepacket $|\Psi(t)\rangle$ prepared by vertical
excitation to $\phi_{I}(\boldsymbol{r};\boldsymbol{R})$:

\begin{equation}
  \sigma_{I}(E) \propto E \int_{-\infty}^{\infty} a_{I}(t) \exp(iEt)
  dt,
\end{equation}

\noindent
with

\begin{equation}
  a_{I}(t) = \left\langle \Psi(0) \middle| \Psi(t) \right\rangle,
\end{equation}

\begin{equation}
  \left| \Psi(0) \right\rangle = \left\{ \left| \phi_{I}
  \right\rangle \left\langle \phi_{1} \right| + h.c. \right\} \left|
  \Psi_{GS} \right\rangle.
\end{equation}

\noindent
Here, $| \Psi_{GS} \rangle$ denotes the ground vibronic state, which
was obtained {\it via} wavepacket
relaxation\cite{kosloff_relaxation}.

All wavepacket propagations were performed using the
multiconfigurational time-dependent Hartree (MCTDH)
approach\cite{mctdh-meyer90,mctdh-review,mctdh_review_2008}, as
implemented in the Quantics quantum dynamics
package\cite{quantics}. The multi-set formalism was used, in which the
wavefunction {\it ansatz} reads

\begin{equation}
  \left| \Psi(\boldsymbol{q},t) \right\rangle = \sum_{I=1} \left|
  \Psi^{(I)} (\boldsymbol{q}, t) \right\rangle \left| \phi_{I}
  \right\rangle.
\end{equation}

\noindent
Here, $| \Psi^{(I)} (\boldsymbol{q}, t) \rangle$ is the nuclear
wavefunction for the $I$th electronic state, which is is expanded in a
direct product basis formed from time-dependent functions
$\varphi_{j_{k}^{I}}^{(\kappa,I)}$, termed single-particle functions
(SPFs):

\begin{equation}
  \left| \Psi^{(I)} (\boldsymbol{q}, t) \right\rangle =
  \sum_{j_{1}^{I}=1}^{n_{1}^{I}} \cdots \sum_{j_{p}^{I}=p}^{n_{p}^{I}}
  A_{j_{1}^{I},\dots,j_{p}^{I}}^{(I)} (t) \prod_{\kappa=1}^{p}
  \varphi_{j_{\kappa}^{I}}^{(\kappa, I)} (q_{\kappa}, t).
\end{equation}

\noindent
The SPFs are functions of generally multi-dimensional logical
coordinates $q_{\kappa}$, each corresponding to a composite of
$d_{\kappa}$ physical nuclear coordinates $R_{\nu}^{(\kappa)}$:

\begin{equation}
  q_{\kappa} = \left( R_{1}^{(\kappa)}, \dots,
    R_{d_{\kappa}}^{(\kappa)} \right).
\end{equation}

\noindent
The time-dependent SPFs are further expanded in terms of a
time-independent discrete variable representation
(DVR)\cite{mctdh-review, light_dvr}. Equations of motion for both the
expansion coefficients
$A_{j_{1}^{I},\dots,j_{p}^{I}}^{(I)}$ and the SPFs are
derived variationally, yielding an optimal description of the evolving
wavepacket\cite{mctdh-review}.

In order to perform the MCTDH wavepacket propagations, the molecular
Hamiltonian is represented in terms of the quasi-diabatic states $\{
\phi_{I}(\boldsymbol{r};\boldsymbol{R}) \}$ furnished by the DFT/MRCI
P-BDD calculations:

\begin{equation}
  \begin{aligned}
    \hat{H} &= \sum_{IJ} \left| \phi_{I}(\boldsymbol{r};
    \boldsymbol{R}) \right\rangle \left\langle
    \phi_{I}(\boldsymbol{r}; \boldsymbol{R}) \right| \hat{H} \left|
    \phi_{J}(\boldsymbol{r}; \boldsymbol{R}) \right\rangle
    \left\langle \phi_{J}(\boldsymbol{r}; \boldsymbol{R}) \right|
    \\ &= \sum_{I} \left| \phi_{I}(\boldsymbol{r}; \boldsymbol{R})
    \right\rangle \hat{T}_{n}(\boldsymbol{R}) \left\langle
    \phi_{I}(\boldsymbol{r}; \boldsymbol{R}) \right|
    \\ &+ \sum_{IJ} \left| \phi_{I}(\boldsymbol{r}; \boldsymbol{R})
    \right\rangle W_{IJ}(\boldsymbol{R}) \left\langle
    \phi_{J}(\boldsymbol{r}; \boldsymbol{R}) \right|.
  \end{aligned}
\end{equation}

\noindent
where the $W_{IJ}(\boldsymbol{R}) = \langle \phi_{I} | \hat{H}_{el} |
\phi_{J} \rangle$ are the elements of the quasi-diabatic potential
matrix $\boldsymbol{W}(\boldsymbol{R})$. To proceed,
$\boldsymbol{W}(\boldsymbol{R})$ must be re-cast in a form that is
compatible with MCTDH. Specifically, as a sum of products of monomodal
operators\cite{mctdh-review}. This is achieved by approximating
$\boldsymbol{W}(\boldsymbol{R})$ using the vibronic coupling
Hamiltonian model of K\"{o}ppel {\it et
  al.}\cite{koppel84,cederbaum-vcham}. Briefly,
$\boldsymbol{W}(\boldsymbol{R})$ is represented by a Taylor expansion
in terms of mass- and frequency-scaled normal modes, $Q_{\alpha}$,
about the ground state minimum $\boldsymbol{Q}_{0}$. Our model
potential is complete up to fourth-order in the one-mode terms and
contains only bi-linear two-mode terms:

\begin{equation}\label{eq:modpot}
  \begin{aligned}
    W_{IJ}(\boldsymbol{R}) &\approx W_{IJ}^{mod}(\boldsymbol{Q}) \\
    &= \tau_{0}^{(I,J)} + \sum_{p=1}^{4} \frac{1}{p!}  \sum_{\alpha}
    \tau_{p\alpha}^{(I,J)} Q_{\alpha}^{p} + \frac{1}{2} \sum_{\alpha
      \beta} \eta_{\alpha \beta}^{(I,J)} Q_{\alpha} Q_{\beta}.
  \end{aligned}
\end{equation}

\noindent
The coupling coefficients $\boldsymbol{\tau}_{p}^{(I,J)}$ and
$\boldsymbol{\eta}^{(I,J)}$ were calculated using a normal equations
approach, as detailed in Appendix \ref{app:fitting}.

We note that, although only approximations to the functions of
interest, if the model potentials used yield accurate absorption
spectra, then it can be concluded that the DFT/MRCI P-BDD potentials
to which they are fitted are of good quality, which is what we aim to
establish.

\subsection{Quantum Chemistry Calculations}
In all DFT/MRCI calculations, the R2016 Hamiltonian was
used\cite{lyskov_dftmrci_redesign}. The parameterisation of this
Hamiltonian was performed using the BH-LYP functional, and,
accordingly, this was used in all calculations. The KS orbitals used
were computed using the Turbomole set of
programs\cite{turbomole_v6.1}.

For the pyrazine calculations, the TZVP basis was used in all DFT/MRCI
calculations. In the construction of the vibronic coupling
Hamiltonian, normal modes and frequencies were calculated at the
MP2/TZVP level of theory. Both the geometry optimisation and frequency
calculations were performed using the Turbomole set of
programs\cite{turbomole_v6.1}.

In the LiH calculations, the QZVPP basis set was used. In addition to
the DFT/MRCI calculations, canonical MRCI calculations were also
performed to provide benchmark derivative coupling values. In these
calculations, a (2,5) active space comprised of the $1s_{H}$,
$2s_{Li}$, and the full set of $2p_{Li}$ orbitals was used. These
calculations were performed using the Columbus set of
programs\cite{columbus}.

\section{Results}\label{sec:results}
We here present the results of P-BDD calculations performed using
DFT/MRCI wavefunctions.

In Section \ref{results:lih}, we discuss the diabatic potentials
calculated for the two lowest-lying $^{1}\Sigma^{+}$ states of LiH
using the combination of the DFT/MRCI and P-BDD methodologies. In
Section \ref{results:pyrazine}, we consider the results of spectral
simulations performed for pyrazine using a model potential derived
from DFT/MRCI P-BDD calculations. We also present an analysis of the
sensitivity of the wavepacket propagation to the errors introduced
into the P-BDD diabatic potentials through the use of different levels
of approximation. Namely, Hadamard screening and wavefunction
truncation.

\subsection{LiH}\label{results:lih}
The $1^{1}\Sigma^{+}$ and $2^{1}\Sigma^{+}$ adiabatic states of LiH
display an ionic-covalent avoided crossing as the Li-H bond is
stretched. At the equilibrium bond length, the $1^{1}\Sigma^{+}$ state
has an ionic $(1s_{Li})^{2}(1s_{H})^{2}$ character, while the
$2^{1}\Sigma^{+}$ state has $(1s_{Li})^{2}(1s_{H})(2s_{Li})$
character. The two states, however, are strongly coupled by the Li-H
stretching coordinate, resulting in a pronounced avoided crossing and
the mixing of the state characters as the bond is elongated.

For clarity, the following state labeling convention shall be
used. The adiabatic states of interest will be labeled by the term
symbols $1^{1}\Sigma^{+}$ and $2^{1}\Sigma^{+}$. The diabatic states
derived from these shall be labeled by $\tilde{X}$ and $\tilde{A}$.

\subsubsection{Adiabatic and Diabatic Potentials}
Using the P-BDD diabatisation procedure, the $1^{1}\Sigma^{+}$ and
$2^{1}\Sigma^{+}$ states were rotated to a diabatic representation. In
the P-BDD calculations, the adiabatic and diabatic representations
were taken to be equal at the $1^{1}\Sigma^{+}$ minimum energy bond
length, denoted by $r_{0}$. Owing to the small size of the problem, no
approximations (wavefunction truncation and Hadamard screening) were
used in the P-BDD calculations.

\begin{figure}
  \begin{center}
    \includegraphics[width=7.0cm,angle=0]{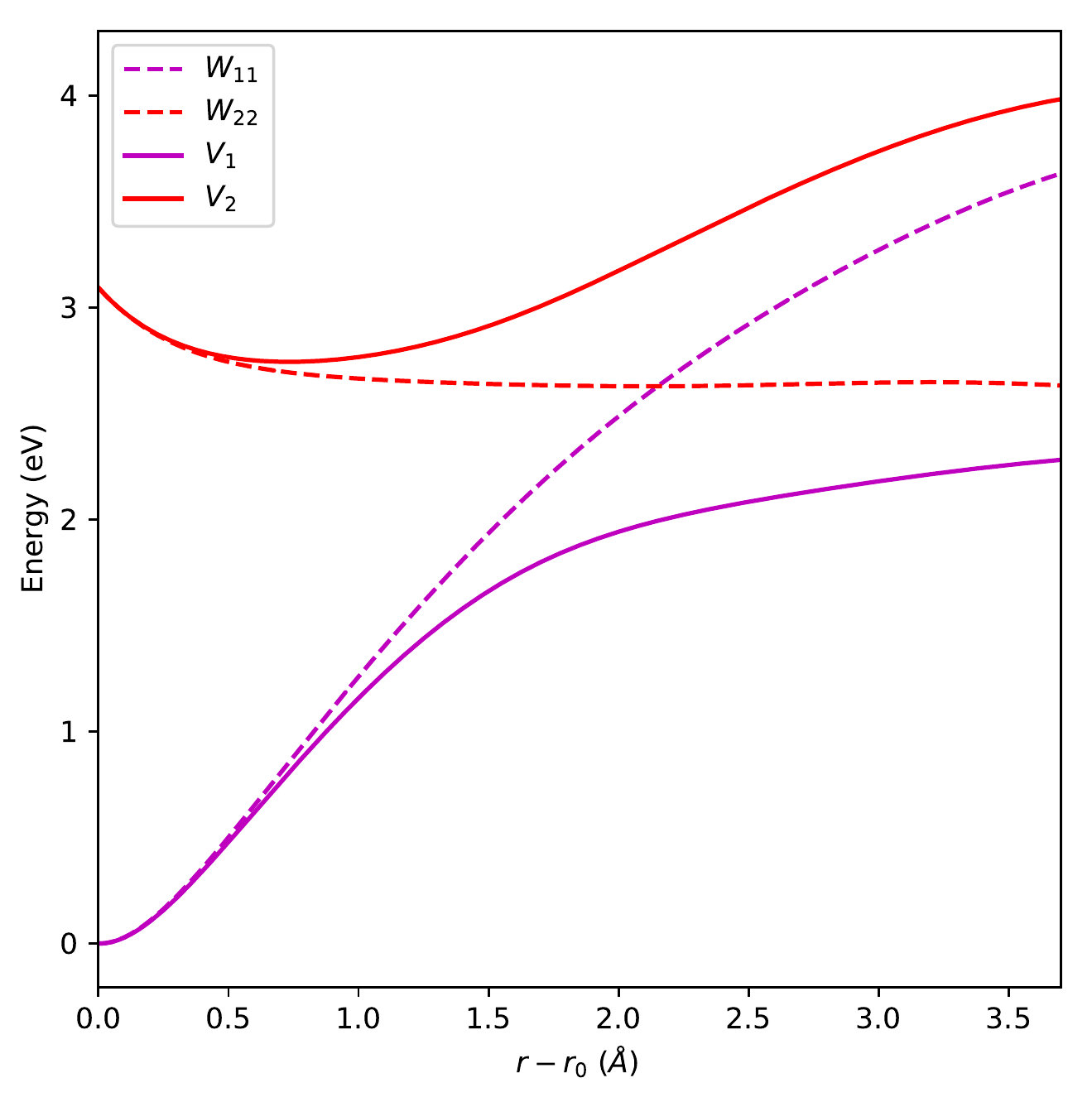}
    \includegraphics[width=7.0cm,angle=0]{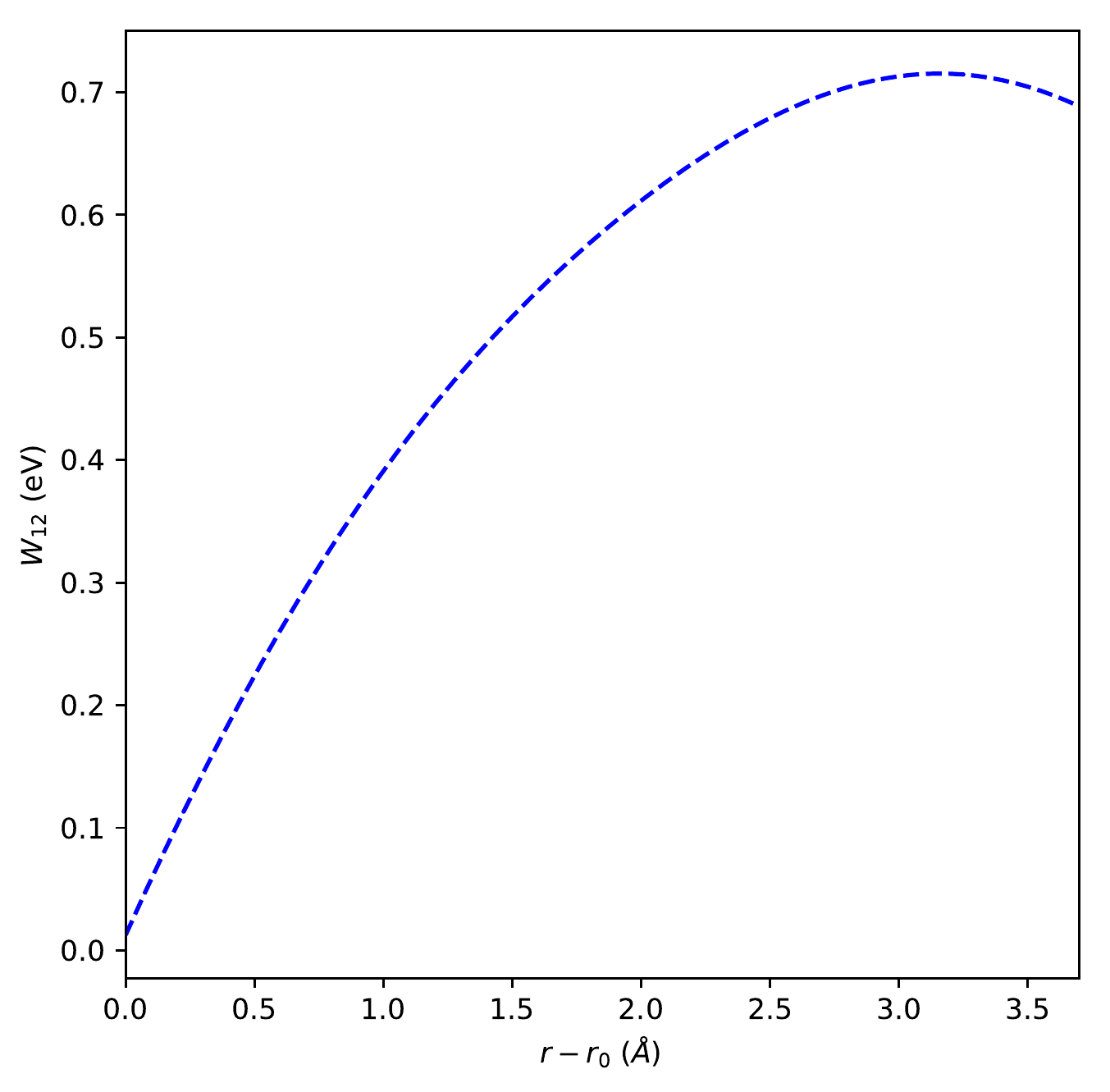}
    \caption{Top: Adiabatic and diabatic potential energy surfaces for
      the first two $^{1}\Sigma^{+}$ states of LiH calculated at the
      DFT/MRCI/QZVPP level of theory. Bottom diabatic coupling between
      the $\tilde{X}$ and $\tilde{A}$ states, also calculated at the
      DFT/MRCI/QZVPP level of theory. All diabatic quantities were
      calculated using the P-BDD method.}
    \label{fig:lih_diabpot}
  \end{center}
\end{figure}

The calculated adiabatic and diabatic potentials are shown in Figure
\ref{fig:lih_diabpot}. Also shown in Figure \ref{fig:lih_diabpot} is
the diabatic coupling, $W_{12}$, between the $\tilde{X}$ and
$\tilde{A}$ states as a function of bond length. The DFT/MRCI P-BDD
diabatic potential is found to capture the strong non-adiabatic
coupling between the $1^{1}\Sigma^{+}$ and $2^{1}\Sigma^{+}$ states,
with the avoided crossing being correctly removed in the diabatic
representation.

\subsubsection{Derivative Couplings}
Further, more conclusive evidence that the DFT/MRCI
P-BDD diabatic potentials correctly account for the non-adiabatic
coupling of the $1^{1}\Sigma^{+}$ and $2^{1}\Sigma^{+}$ states comes
from a consideration of the derivative coupling, $F_{12}(r)$, between
the two states as derived from the DFT/MRCI P-BDD diabatic potential
$\boldsymbol{W}(r)$,

\begin{equation}
    F_{12}(r) = \left\langle \psi_{1}(r) \middle|
    \frac{\partial}{\partial r} \middle| \psi_{2}(r) \right\rangle =
    \frac{\left[ \boldsymbol{U}(r) \frac{\partial
          \boldsymbol{W}(r)}{\partial r} \boldsymbol{U}^{\dagger}(r)
        \right]_{12}}{V_{2}(r)-V_{1}(r)},
\end{equation}

\noindent
where $\boldsymbol{U}$ denotes the P-BDD ADT matrix and $V_{I}$ its
eigenvalues. To determine the derivatives of the diabatic potential
matrix elements $W_{IJ}$, the calculated values were first fitted to a
tenth-order Chebyshev expansion that was then analytically
differentiated. To provide a benchmark to compare to, derivative
couplings were also calculated analytically at the canonical MRCI
level of theory.

\begin{figure}
  \begin{center}
    \includegraphics[width=7.0cm,angle=0]{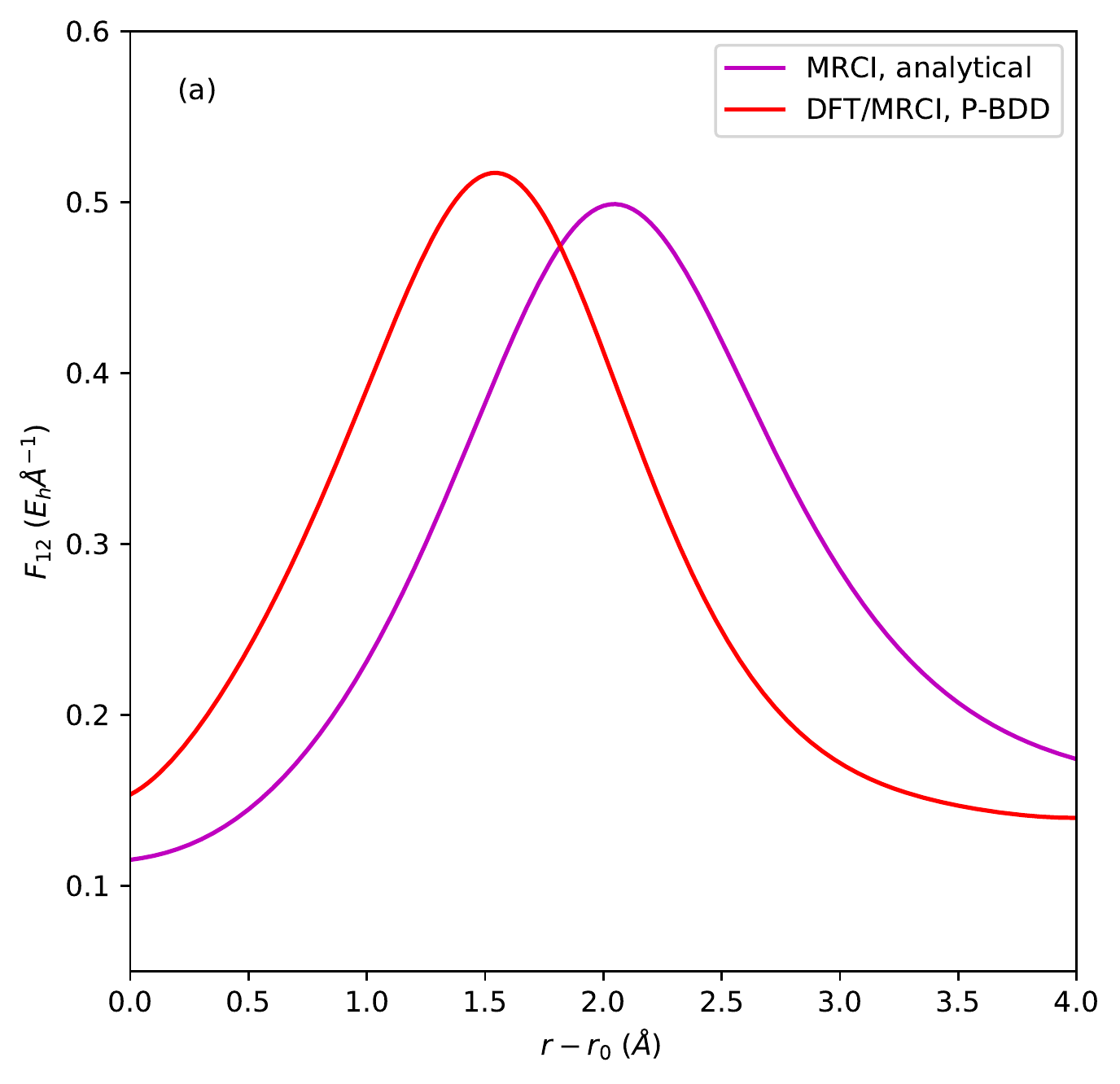}
    \includegraphics[width=7.0cm,angle=0]{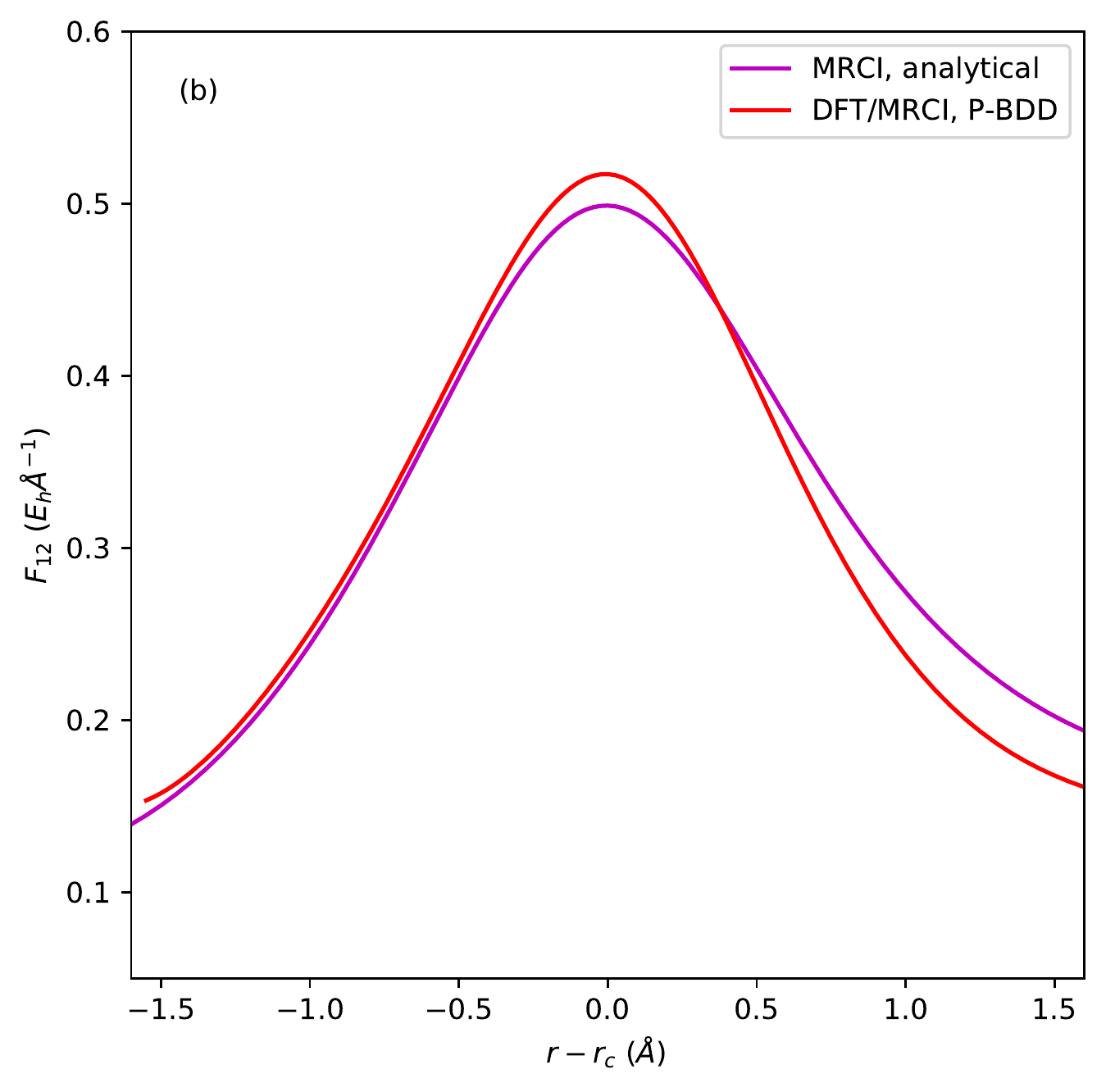}
    \caption{Comparison of derivative coupling terms $F_{12}(r)$
      between the first two $^{1}\Sigma^{+}$ states of LiH calculated
      from the DFT/MRCI P-BDD diabatic potential matrix and
      analytically at the MRCI(2,6) level of theory. (a) Comparison of
      $F_{12}$ values as a function of Li-H bond length. (b)
      Comparison of $F_{12}$ values as a function of the displacement
      from the centre of the avoided crossing between the
      $1^{1}\Sigma^{+}$ and $2^{1}\Sigma^{+}$ states, $r_{c}$. The
      QZVPP basis was used in all calculations.}
    \label{fig:lih_f12}
  \end{center}
\end{figure}

The DFT/MRCI P-BDD and canonical MRCI derivative couplings are shown
alongside each other in Figure \ref{fig:lih_f12}. Importantly, it is
found that the DFT/MRCI P-BDD derivative couplings behave similarly to
the canonical MRCI values as a function of Li-H bond length. In
particular, the maximum value reached (0.51 $E_{h}\AA^{-1}$) compares
very agreeably with the maximum value of the canonical MRCI derivative
couplings (0.49 $E_{h}\AA^{-1}$). The one slight discrepancy between
the two results is that the DFT/MRCI P-BDD values peak around 0.5
$\AA$ before the MRCI values, a result of the DFT/MRCI method yielding
an avoided crossing between the $1^{1}\Sigma^{+}$ and
$2^{1}\Sigma^{+}$ states that is slightly too early compared to the
MRCI result. If, however, the DFT/MRCI P-BDD and canonical MRCI
derivative couplings are plotted as a function of displacement from
the centre of the avoided crossing, $r_{c}$, then almost quantitative
agreement is found, as shown in Figure \ref{fig:lih_f12}.

\subsubsection{Natural Orbital Analysis of Ionic and Covalent State
  Characters}
Finally, we consider the characters of the DFT/MRCI P-BDD diabatic
states as a function of the Li-H bond length. Unlike in the adiabatic
representation, the $\tilde{X}$ and $\tilde{A}$ states should maintain
ionic and covalent characters, respectively, as the Li-H bond is
stretched (so long as the bond length $r$ is within the non-asymptotic
region of the potential). To analyse the electronic characters of the
diabatic states at a given bond length, we consider the dominant
natural orbitals (NOs) derived from the density matrices constructed
using the DFT/MRCI P-BDD diabatic wavefunctions. Remembering that the
electronic density $\rho(\boldsymbol{r})$ for a given state can be
expressed in terms of NOs $\varphi_{p}$ and natural occupations
$\lambda_{p}$ as

\begin{equation}
  \rho(\boldsymbol{r}) = \sum_{p} \lambda_{p} \left| \varphi_{p}
  \right|^{2},
\end{equation}

\noindent
the spatial (de)localisation of the dominant NOs can be used to assess
the degree of ionic versus covalent character of a given electronic
state.

\begin{figure}
  \begin{center}
    \includegraphics[width=4.0cm,angle=0]{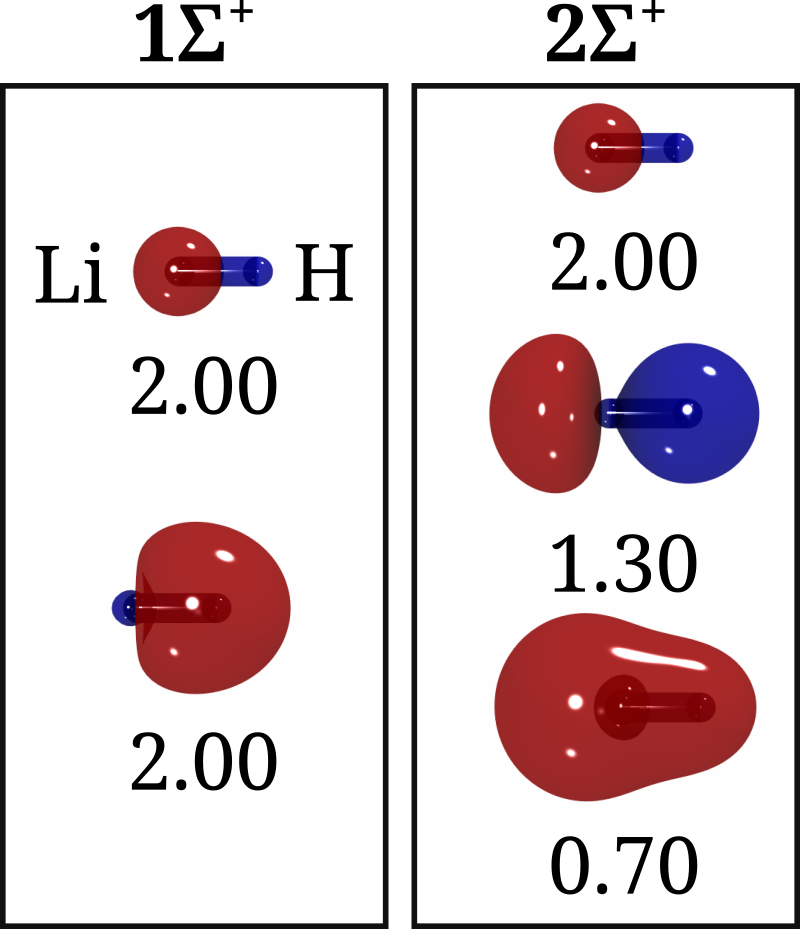}
    \caption{Natural orbitals for $1^{1}\Sigma^{+}$ and
      $2^{1}\Sigma^{+}$ states at the equilibrium bond length,
      $r_{0}$, calculated at the DFT/MRCI/QZVPP level of theory. The
      numbers below each natural orbital is the corresponding natural
      occupation. For the first NO for the $1^{1}\Sigma^{+}$ state,
      the Li and H atom labels are also given.}
    \label{fig:lih_r0_nos}
  \end{center}
\end{figure}

\begin{figure}
  \begin{center}
    \includegraphics[width=7.0cm,angle=0]{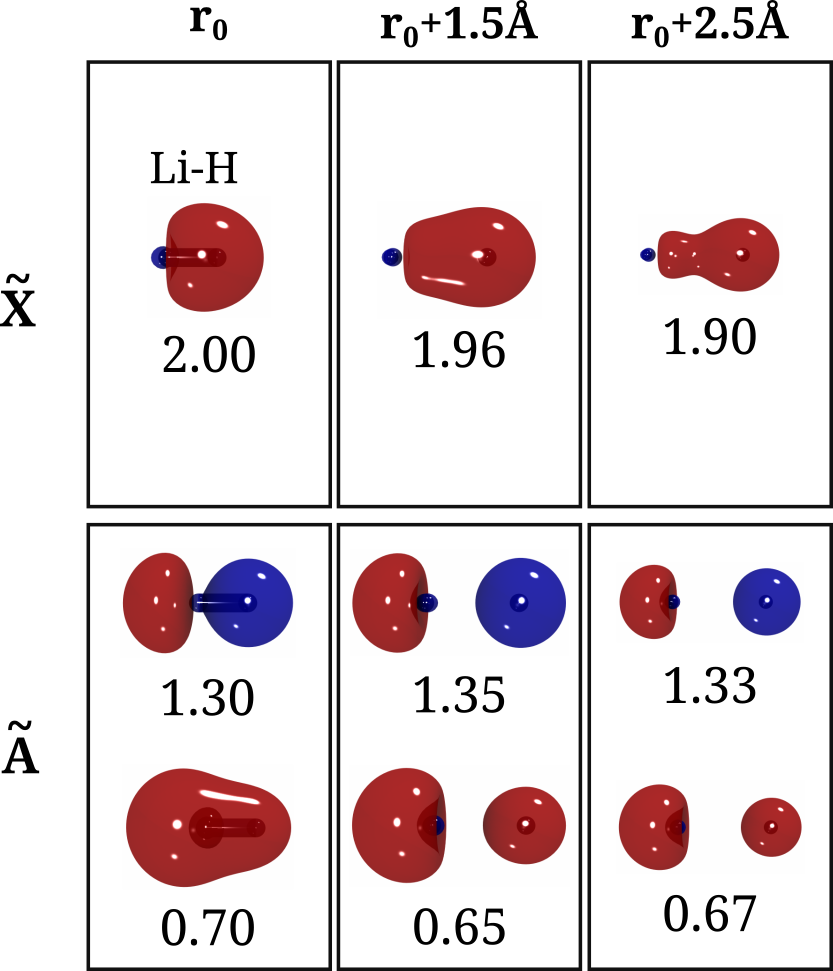}
    \caption{Diabatic natural orbitals for $\tilde{X}$ and $\tilde{A}$
      states of LiH as a function of bond length calculated at the
      DFT/MRCI/QZVPP P-BDD level of theory. Here, $r_{0}$ denotes the
      equilibrium bond length. The numbers below each natural orbital
      is the corresponding natural occupation. Excluded are the NOs
      corresponding to the $1s_{Li}$ orbital.}
    \label{fig:lih_diab_nos}
  \end{center}
\end{figure}

\begin{figure}
  \begin{center}
    \includegraphics[width=7.0cm,angle=0]{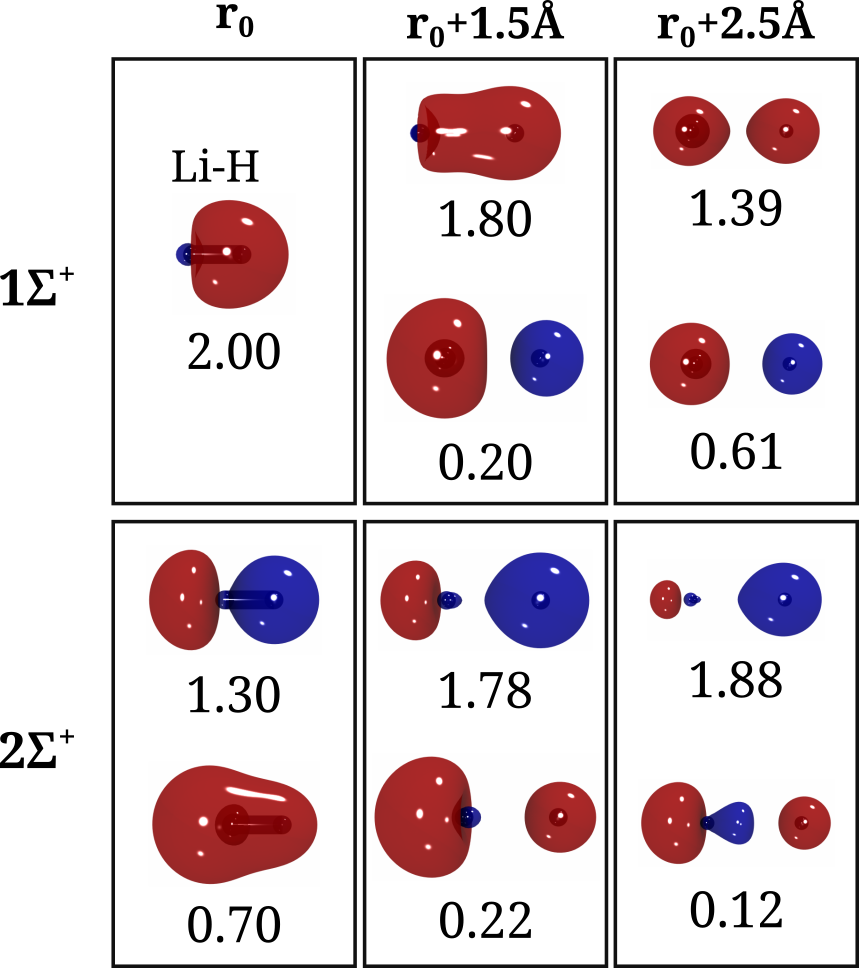}
    \caption{Adiabatic natural orbitals for $1^{1}\Sigma^{+}$ and
      $2^{1}\Sigma^{+}$ states of LiH as a function of bond length
      calculated at the DFT/MRCI/QZVPP level of theory. Here, $r_{0}$
      denotes the equilibrium bond length. The numbers below each
      natural orbital is the corresponding natural
      occupation. Excluded are the NOs corresponding to the $1s_{Li}$
      orbital.}
    \label{fig:lih_adiab_nos}
  \end{center}
\end{figure}

We first consider the $1^{1}\Sigma^{+}$ and $2^{1}\Sigma^{+}$ state
NOs at the equilibrium bond length, $r_{0}$, where these states have,
respectively, ionic and covalent characters. These are shown in Figure
\ref{fig:lih_r0_nos}. The ionic character of the $1^{1}\Sigma^{+}$
state at $r_{0}$ is clearly reflected in the NOs: only two NOs exist
with non-zero natural occupations, corresponding each to a $1s_{Li}$
and $1s_{H}$ type orbital. Both have natural occupations of 2.0. This
corresponds, loosely speaking, to a net partial positive charge
associated with the Li-atom and a net partial negative charge
associated with the H-atom (owing to their different nuclear
charges). On the other hand, the NOs for the $2^{1}\Sigma^{+}$ state
show an increase in the localisation of the electron density over the
Li-atom, with the doubly-occupied $1s_{H}$ NO being replaced with two
fractionally-occupied NOs corresponding to linear combinations of the
$2s_{Li}$ and $1s_{H}$ orbitals. This can be clearly be seen to
correspond to a more valence-type electronic character.

Next, we consider the NOs for the $\tilde{X}$ and $\tilde{A}$ states
as a function of the Li-H bond length. The dominant NOs at different
L-H bond lengths are shown in Figure \ref{fig:lih_diab_nos} along with
the corresponding natural occupations. As the occupied NO
corresponding to the $1s_{Li}$ orbital remains virtually unchanged in
character and occupation, it is omitted. At the equilibrium bond
length, $r_{0}$, the adiabatic and diabatic representations are equal,
yielding $\tilde{X}$ and $\tilde{A}$ states with ionic and covalent
character, respectively. From the NOs at elongated Li-H bond lengths
of $r_{0}$+1.5 $\AA$ and $r_{0}$+2.5 $\AA$, it can clearly be seen
that the $r_{0}$ ionic and covalent characters of the $\tilde{X}$ and
$\tilde{A}$ states are preserved, as they should be for good diabatic
states. In contrast the NOs for the $1^{1}\Sigma^{+}$ and
$2^{1}\Sigma^{+}$ states (shown in Figure \ref{fig:lih_adiab_nos})
show that the pure ionic and covalent characters of the adiabatic
states is lost, as a result of their non-adiabatic coupling, as the
Li-H bond is elongated.  Particularly striking is loss of ionic
character of the $1^{1}\Sigma^{+}$ adiabatic state: at $r_{0}$+1.5
$\AA$, the dominant (non-$1s_{Li}$) NOs clearly have non-zero values
around the Li atom, and at $r_{0}$+2.5 $\AA$ the $2^{1}\Sigma^{+}$ has
developed essentially pure covalent character.

\subsection{Pyrazine Absorption Spectrum
  Simulation}\label{results:pyrazine}

\begin{table}
  \caption{Franck-Condon point vertical excitation energies, $\Delta
    E$, and oscillator strengths, $f$, for the first three excited
    states of pyrazine calculated at the DFT/MRCI level of theory
    using the TZVP basis. All energies are given in units of eV.}
  \centering
  \begin{tabular}{lll}
    \hline\hline
    State & $\Delta E$ & $f$ \\
    \hline
    $1B_{3u}(n \pi^{*})$   & 4.09 & 0.012 \\
    $1B_{2u}(\pi \pi^{*})$ & 4.87 & 0.116 \\
    $1A_{u}(n \pi^{*})$    & 5.05 & 0.000 \\
    \hline\hline
  \end{tabular}
  \label{table:pyrazine_vees}
\end{table}

A total of three electronic states were included in the P-BDD
calculations: the $1B_{3u}(n\pi^{*})$, $1B_{2u}(\pi\pi^{*})$ and
$1A_{u}(n\pi^{*})$ states. The Franck-Condon point vertical excitation
energies and oscillator strengths of these states at the DFT/MRCI/TZVP
level of theory are given in Table \ref{table:pyrazine_vees}. To
provide direct comparison with previous studies of pyrazine's
absorption spectrum, we adopt the commonly used normal mode
nomenclature of Innes {\it et al.}\cite{innes_azobenzenes}.

\begin{table}
  \caption{Combined modes, and primitive and SPF basis dimensions used
    in the MCTDH simulations of the pyrazine $1B_{2u}(\pi\pi^{*})$
    state absorption spectrum. For all modes, a harmonic oscillator
    DVR was used as the primitive basis. The numbers $N_{i}$ are
    primitive basis sizes for each physical coordinate. The numbers
    $n_{1},n_{2},n_{3}$ are the SPF basis sizes for each combined mode
    in the $1B_{3u}(n\pi^{*})$, $1B_{2u}(\pi\pi^{*})$ and
    $1A_{u}(n\pi^{*})$ states, respectively.}
  \centering
  \begin{tabular}{lll}
    \hline\hline
    Mode & $N_{i},N_{j}$ & $n_{1},n_{2},n_{3}$ \\
    \hline
    $Q_{2},Q_{6a}$   & 10,48 & 15,12,12 \\
    $Q_{1}$         & 32    & 12,12,12 \\
    $Q_{8a},Q_{9a}$  & 36,22 & 14,12,12 \\
    $Q_{8b},Q_{10a}$ & 22,22 & 12,12,12 \\
    \hline\hline
  \end{tabular}
  \label{table:mctdh_basis}
\end{table}

\begin{figure*}
  \begin{center}
    \includegraphics[width=5.0cm,angle=0]{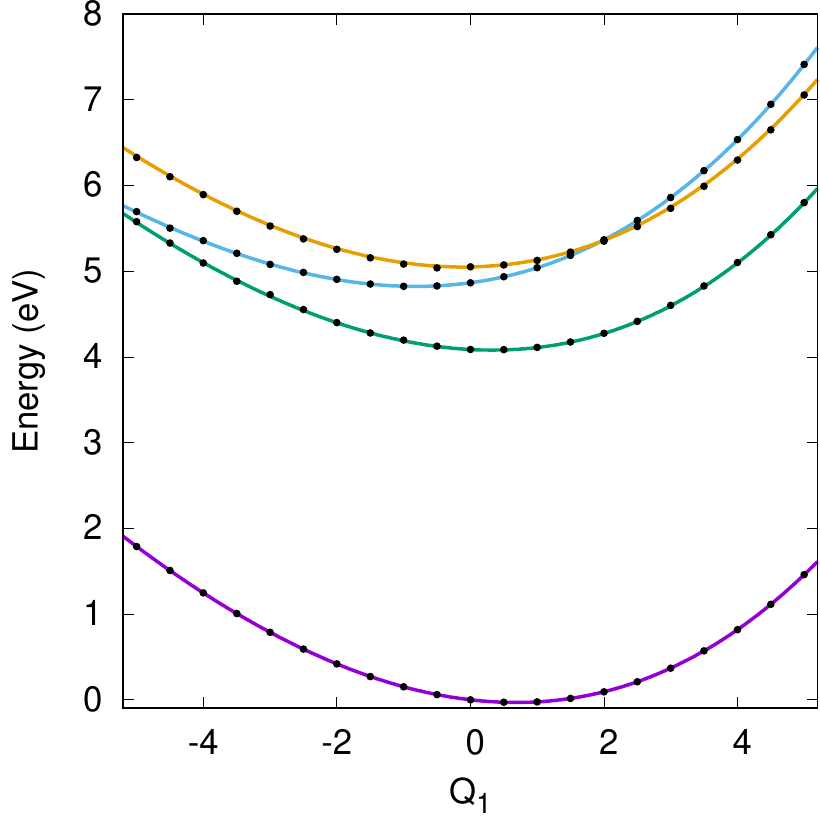}
    \includegraphics[width=5.0cm,angle=0]{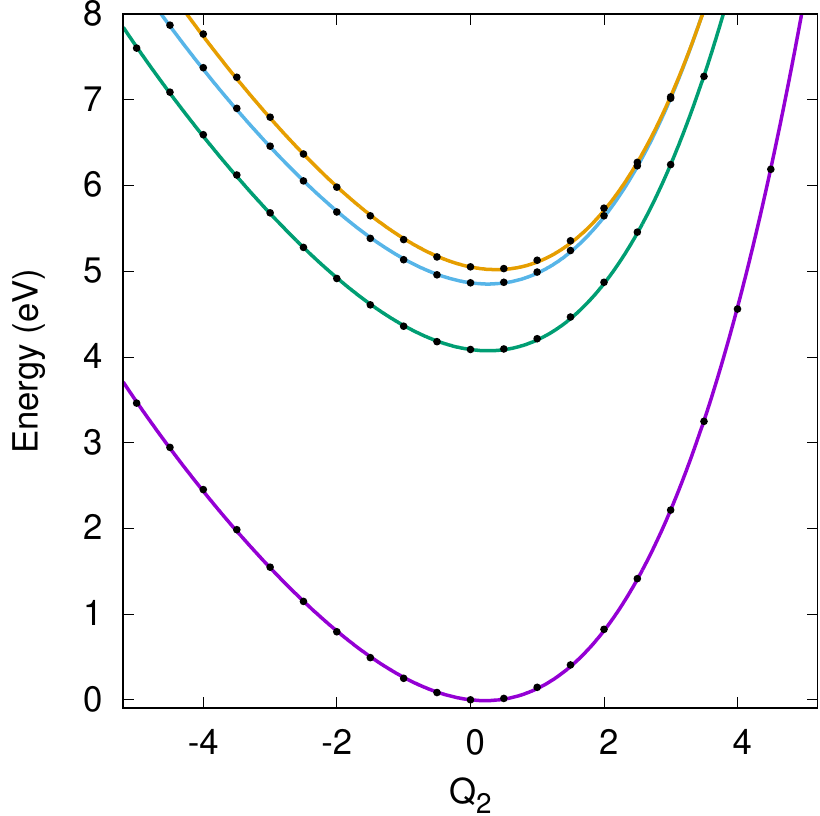}
    \includegraphics[width=5.0cm,angle=0]{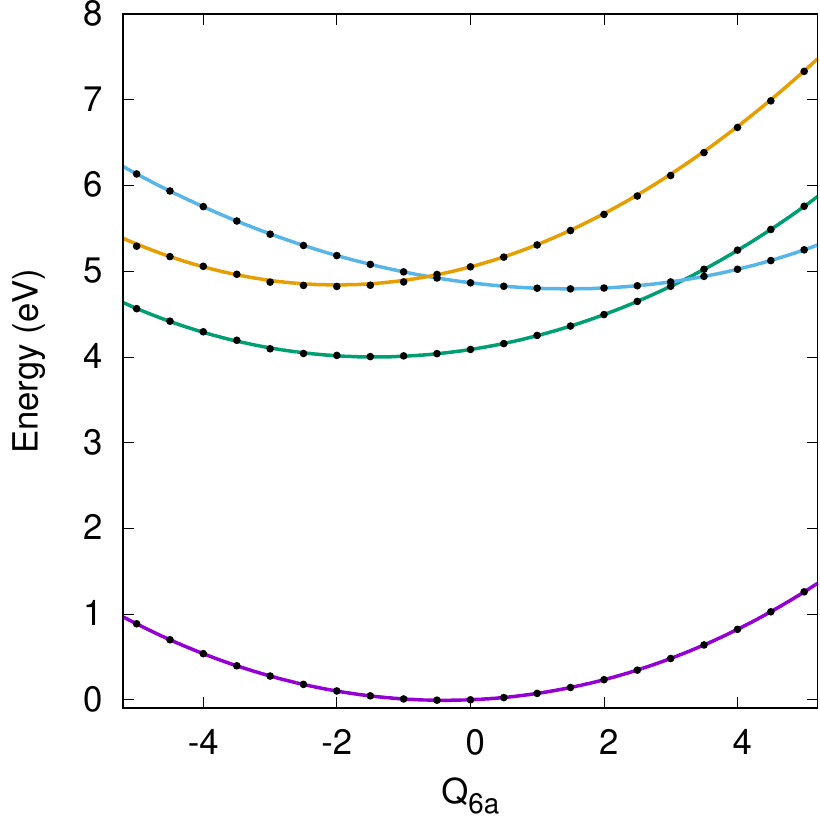}
    \includegraphics[width=5.0cm,angle=0]{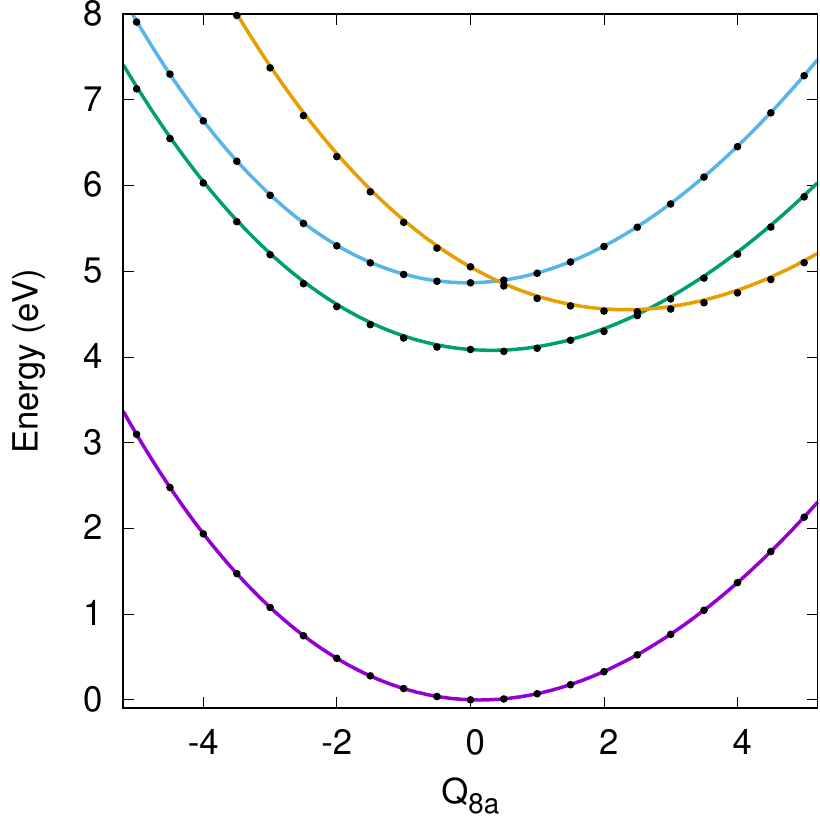}
    \includegraphics[width=5.0cm,angle=0]{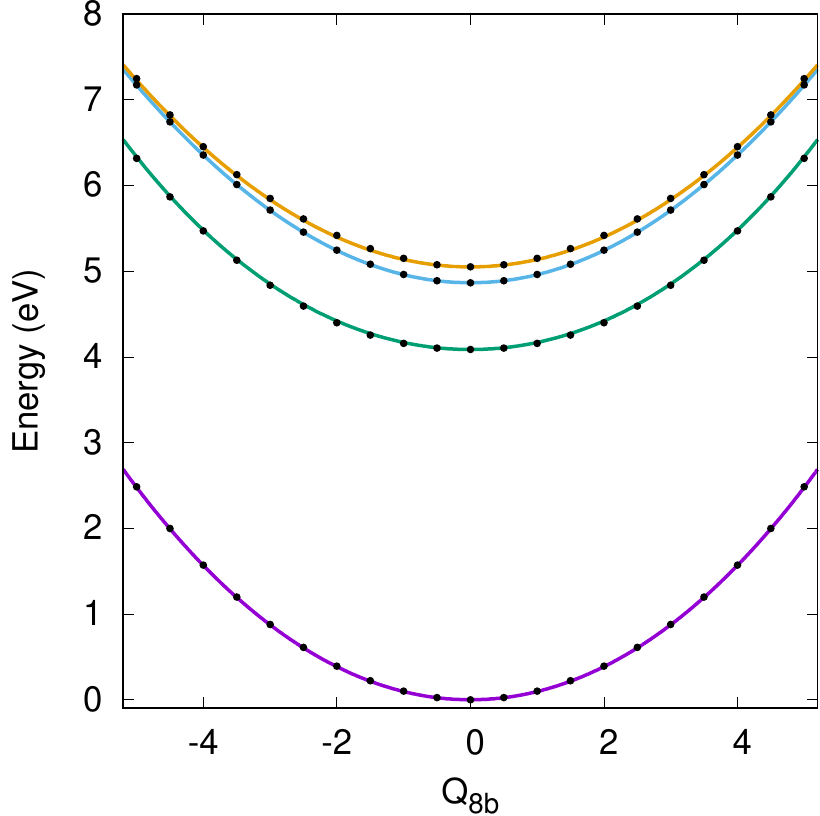}
    \includegraphics[width=5.0cm,angle=0]{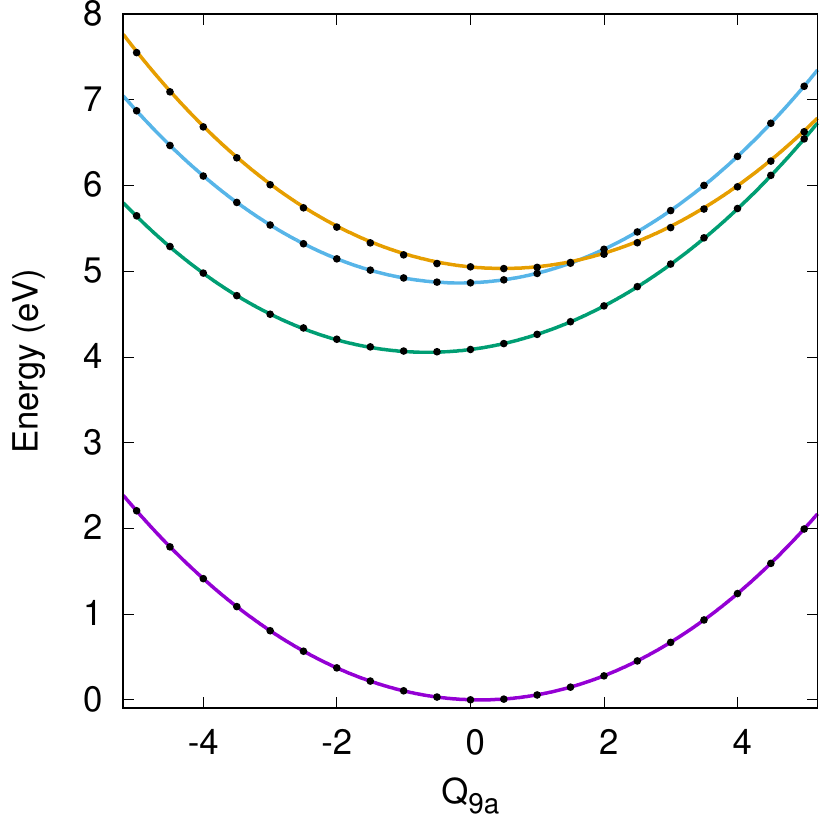}
    \includegraphics[width=5.0cm,angle=0]{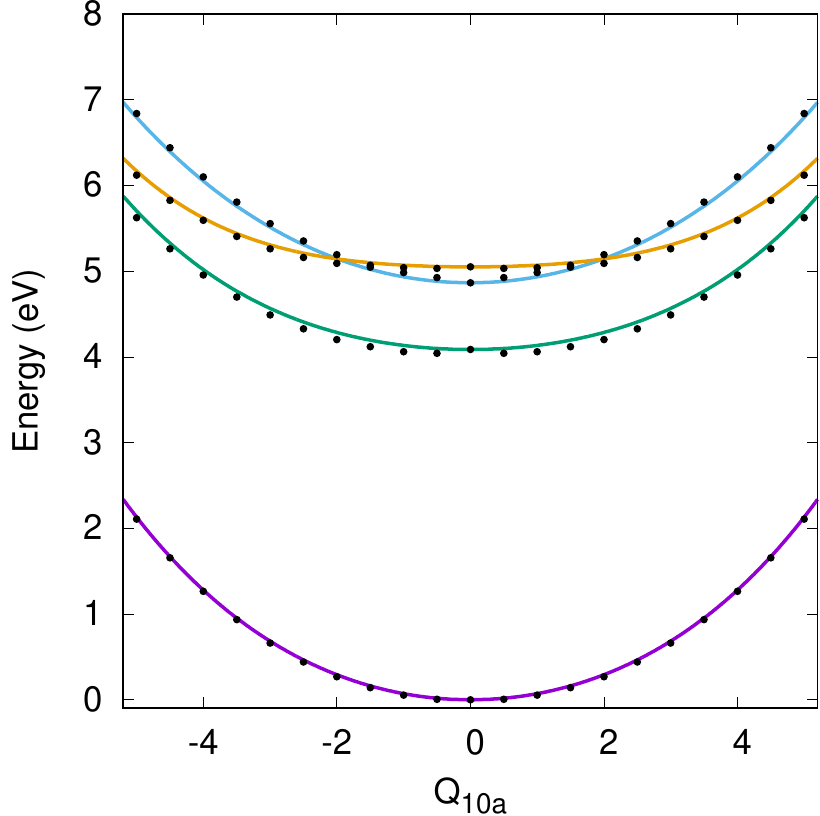}
    \caption{P-BDD diabatic potentials calculated at the DFT/MRCI/TZVP
      level of theory for the first four electronic states in
      pyrazine. The calculated potential values are given by the black
      dots, whilst the solid lines give the model potential values.}
    \label{fig:pyrazine_pots}
  \end{center}
\end{figure*}

In the MCTDH calculations, a propagation time of 120 fs was used,
yielding the wavepacket autocorrelation function for 240
fs\cite{engel_autocorrelation_functions}. The primitive and SPF basis
information is given in Table \ref{table:mctdh_basis}. Seven modes
were included. These are the five totally symmetric ($a_{g}$) modes
$Q_{1}$, $Q_{2}$, $Q_{6a}$, $Q_{8a}$, and $Q_{9a}$. Additionally, the
coupling modes $Q_{8b}$ and $Q_{10a}$ were also included. The mode
$Q_{10a}$ strongly couples the $1B_{3u}(n\pi^{*})$ and
$1B_{2u}(\pi\pi^{*})$ states, whilst the $Q_{8b}$ mode is responsible
for significant coupling of the $1B_{3u}(n\pi^{*})$ and
$1A_{u}(n\pi^{*})$ states. For reference, we show the calculated
DFT/MRCI P-BDD diabatic potentials along these modes in Figure
\ref{fig:pyrazine_pots}. Also shown in here are the model potentials
fitted to the DFT/MRCI P-BDD diabatic potentials. Excellent agreement
between the calculated and model potential values is found along these
cuts. We can thus state with confidence that the accuracy of the
absorption spectrum calculated using the model diabatic potential
$\boldsymbol{W}^{mod}(\boldsymbol{Q})$ can be used to assess the
quality of the DFT/MRCI P-BDD diabatic potentials to which it is was
fitted.

\begin{figure}
  \begin{center}
    \includegraphics[width=7.0cm,angle=0]{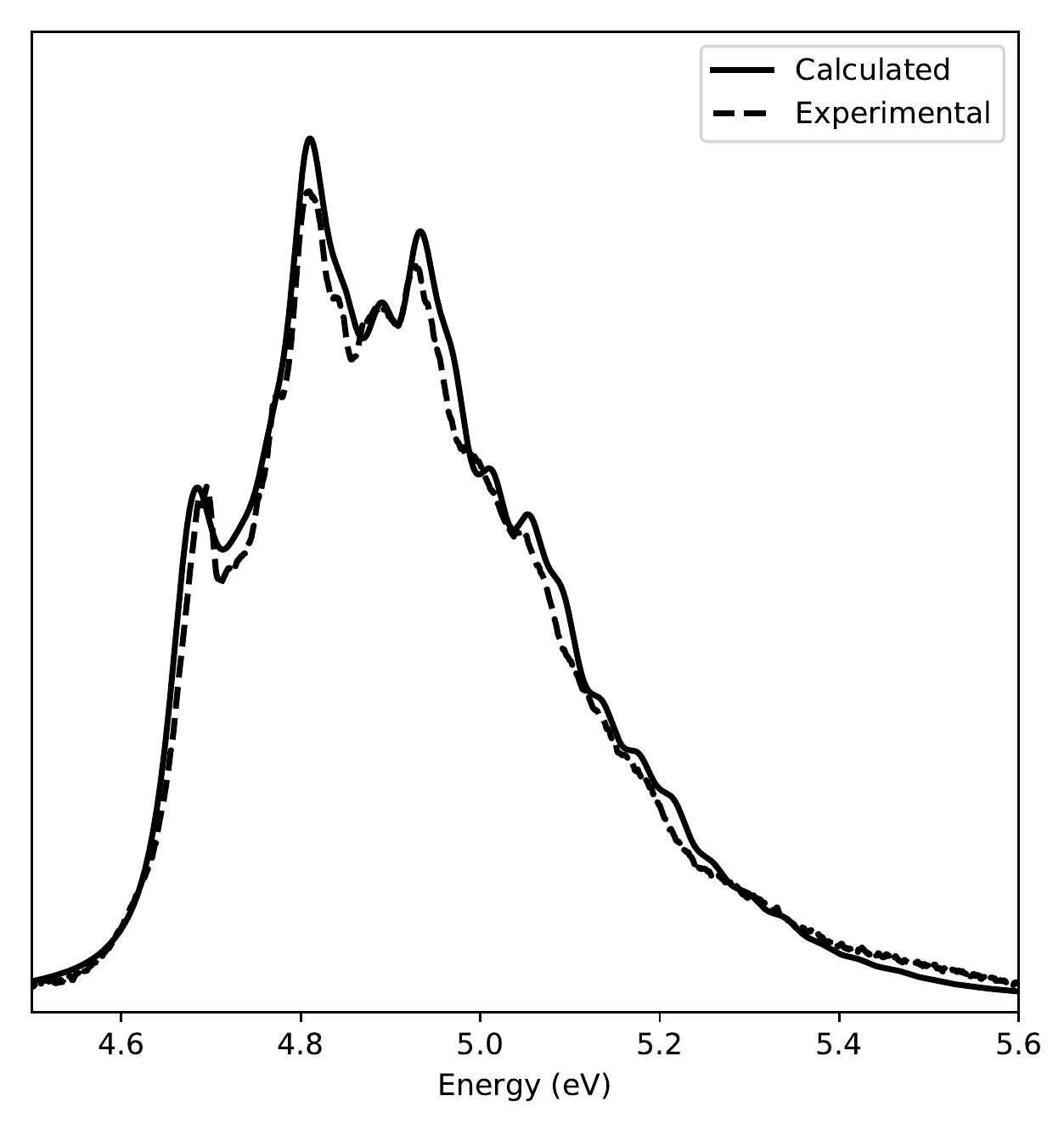}
    \caption{Absorption spectrum corresponding to vertical excitation
      to the $1B_{2u}(\pi\pi^{*})$ state of pyrazine calculated using
      the four-state, seven-mode model potential fitted to diabatic
      potentials calculated using DFT/MRCI and P-BDD with the TZVP
      basis. Shown alongside for comparison is the experimental
      spectrum\cite{shaw_pyrazine_pyrimidine_spectrum}.}
    \label{fig:pyrazine_spec}
  \end{center}
\end{figure}

Shown in Figure \ref{fig:pyrazine_spec} is the absorption spectrum
calculated following vertical excitation to the bright
$1B_{2u}(\pi\pi^{*})$ state. For comparison, we also show the
experimental spectrum of Reference
\citenum{shaw_pyrazine_pyrimidine_spectrum}, taken from the Mainz
spectral database\cite{mainz_spectral_database}. To account for
homogeneous broadening present in the experimental spectrum, the
calculated spectrum was convoluted with a Lorentzian lineshape with a
full width at half maximum (FWHM) of 40 meV. Overall, excellent
agreement between the calculated and experimental spectra is
found. This both validates the model diabatic potential used in its
simulation, as well as the underlying DFT/MRCI P-BDD
calculations. This is an important result, as in order to reproduce
the experimental spectrum, the non-adiabatic coupling between the
$1B_{3u}(n\pi^{*})$ and $1B_{2u}(\pi\pi^{*})$ states must be correctly
described\cite{domcke_pyrazine_1992,werner_pyrazine_1994,worth_pyrazine,raab-pyrazine}.

\subsection{Approximations in the P-BDD
  calculations}\label{sec:approx_p-bdd}
\begin{figure}
  \begin{center}
    \includegraphics[width=7.0cm,angle=0]{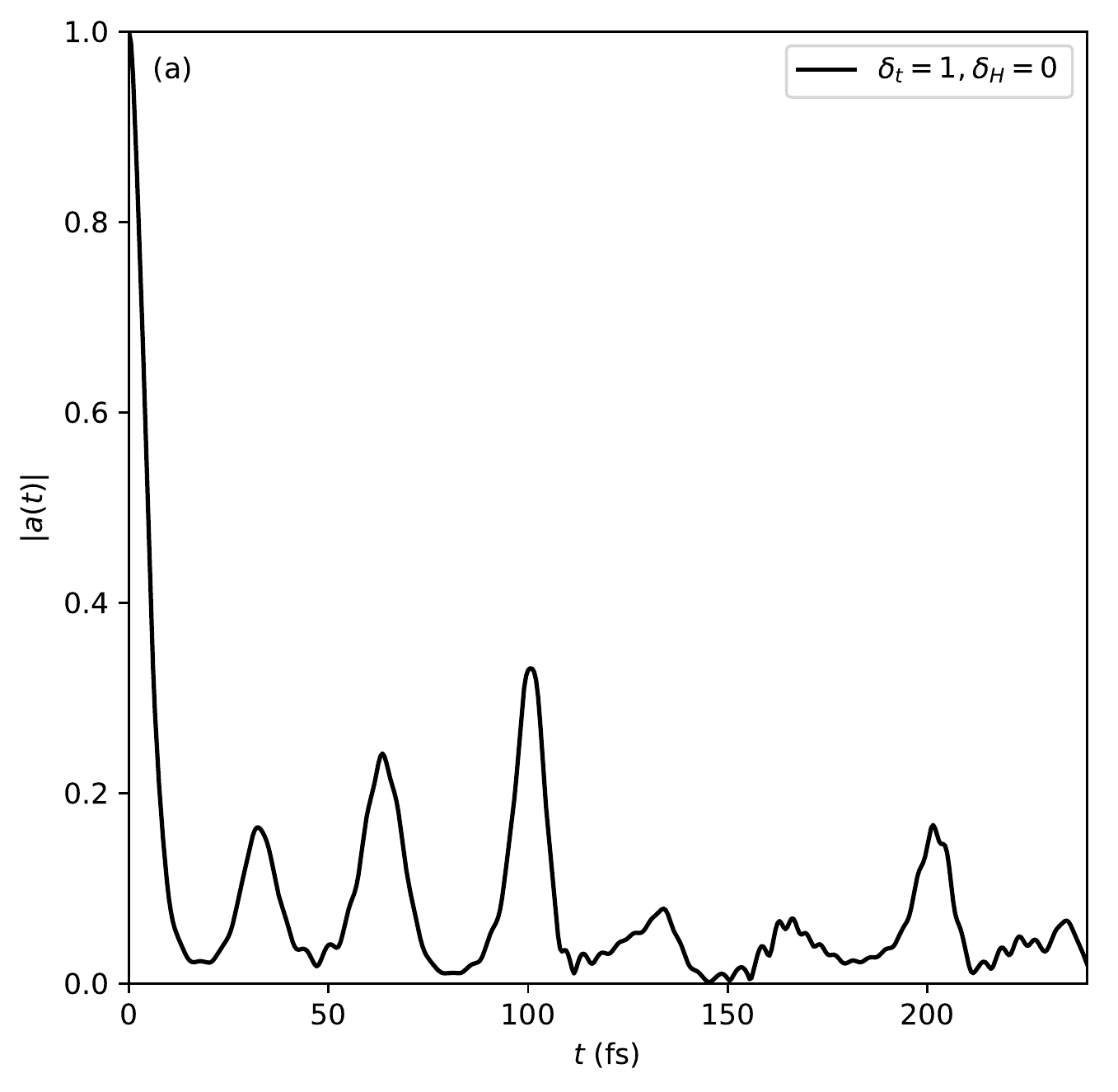}
    \includegraphics[width=7.0cm,angle=0]{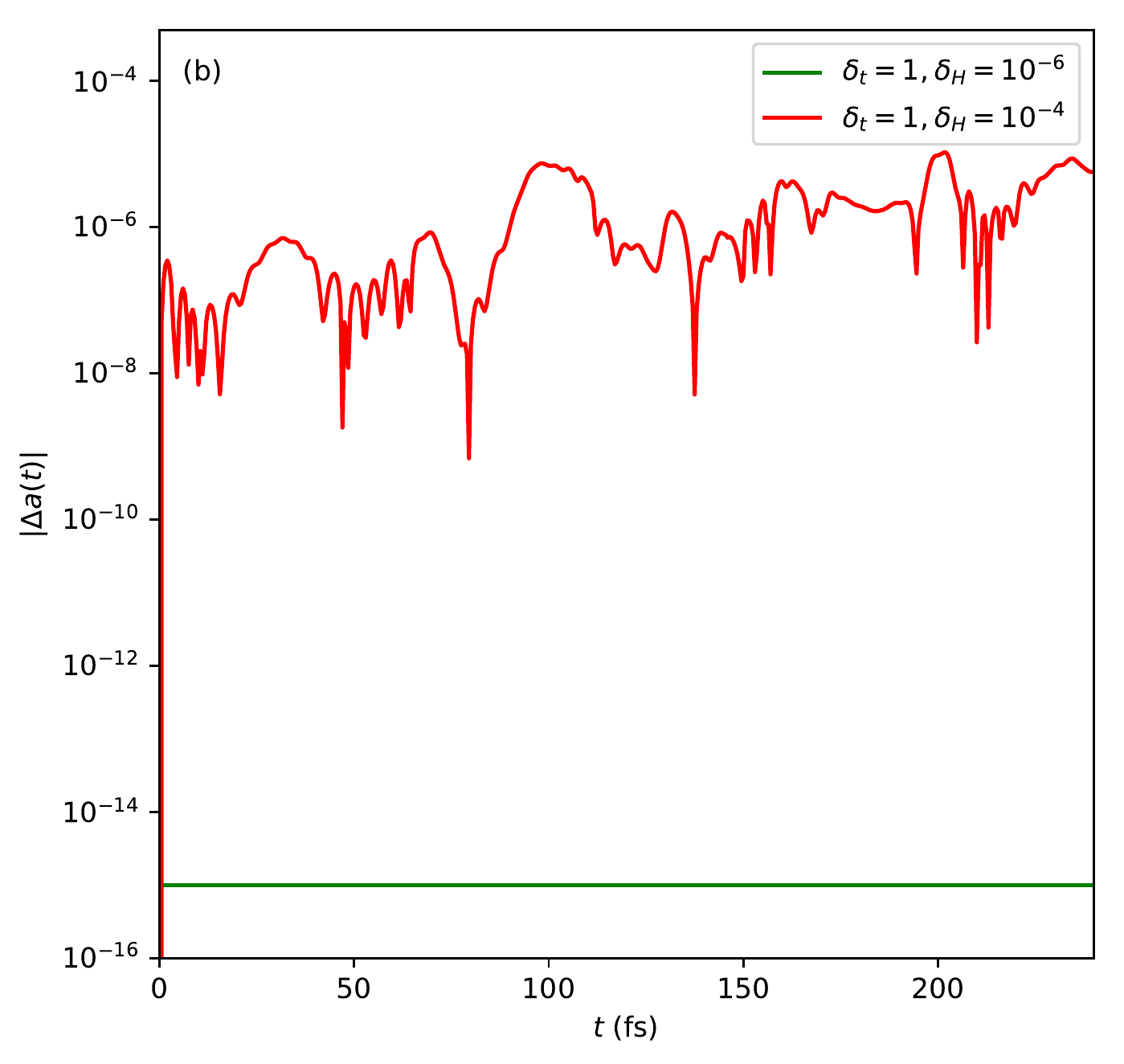}
    \includegraphics[width=7.0cm,angle=0]{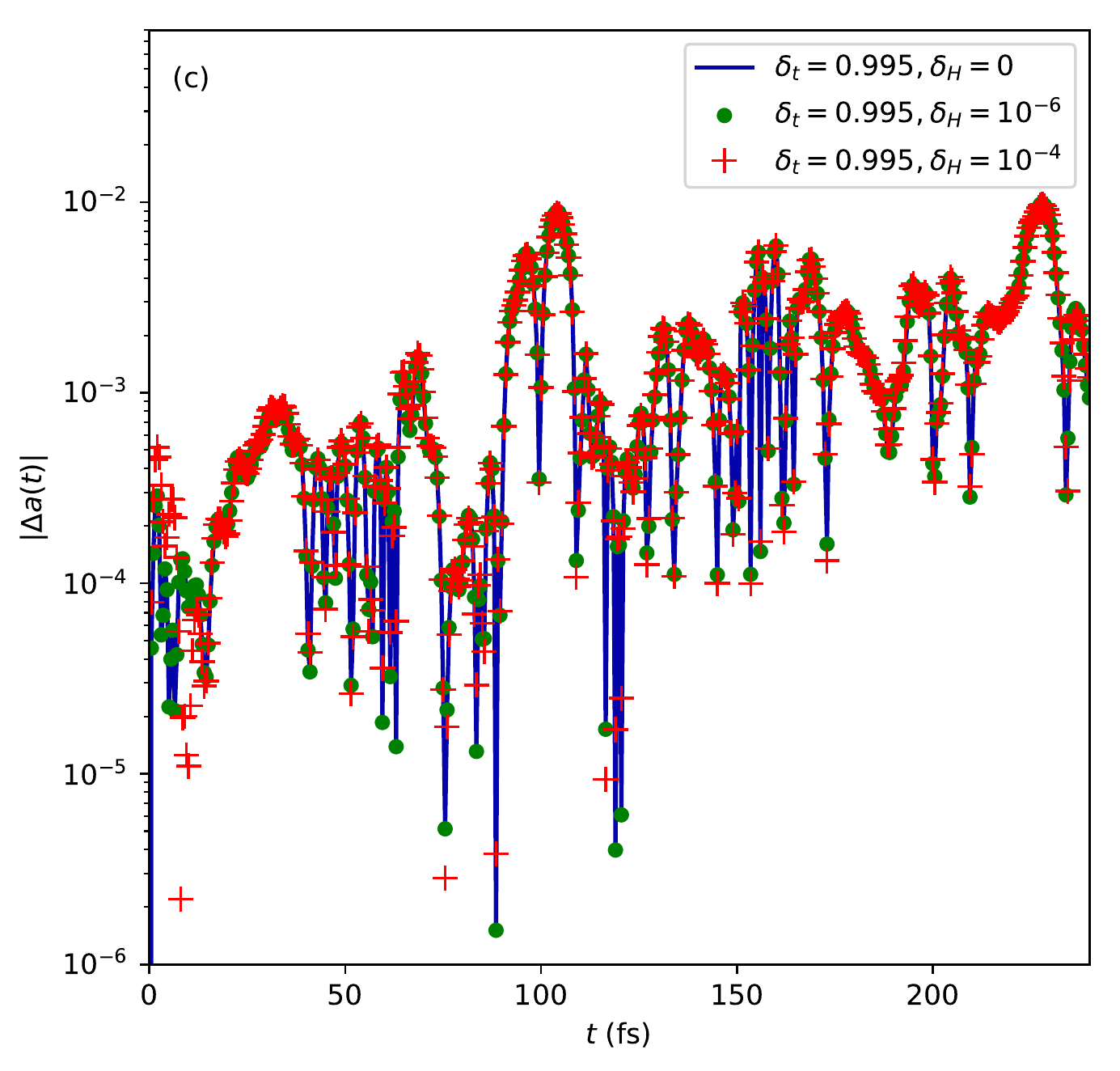}
    \caption{Autocorrelation functions calculated following vertical
      excitation to the $1B_{2u}(\pi\pi^{*})$ state of pyrazine using
      the four-state, seven-mode model potential fitted to diabatic
      potentials calculated using DFT/MRCI and P-BDD. (a)
      Autocorrelation function obtained from potentials calculated
      using P-BDD with exact wavefunction overlaps. (b) Differences in
      the autocorrelation functions calculated using P-BDD with
      different Hadamard screening thresholds, $\delta_{H}$, and
      untruncated wavefunctions. (c) Differences in the
      autocorrelation functions calculated using P-BDD with different
      Hadamard screening thresholds, $\delta_{H}$, and a norm
      truncation threshold $\delta_{t}=0.995$.}
    \label{fig:pyrazine_auto}
  \end{center}
\end{figure}

Having established that the combination of DFT/MRCI and P-BDD is
capable of yielding accurate diabatic potentials, we now consider the
effects of various levels of approximation that may be made in the
P-BDD procedure in order to decrease computational costs. Given two
sets of neighboring adiabatic wavefunctions $\{
\psi_{I}(\boldsymbol{r};\boldsymbol{R}_{n}) \}$ and $\{
\psi_{I}(\boldsymbol{r};\boldsymbol{R}_{n+1}) \}$, the bottleneck in a
P-BDD calculation is in the computation of all overlaps between the
members of the two sets. Here, two approximations may be made: (i)
Hadamard screening of the unique factors $\det(\boldsymbol{s}_{kl})$
and $\det( \bar{\boldsymbol{s}}_{kl})$, and; (ii) the truncation of
the Slater determinant expansions of the wavefunctions
$\psi_{I}(\boldsymbol{r}; \boldsymbol{R})$. We are ultimately
interested in the use of DFT/MRCI P-BDD diabatic potentials in quantum
dynamics simulations. Therefore, we consider the changes in the
wavepacket autocorrelation function $a(t)$ in the pyrazine MCTDH
calculation performed using model potentials derived from DFT/MRCI
P-BDD calculations using different Hadamard screening thresholds and
wavefunction truncations. For reference, we show in Figure
\ref{fig:pyrazine_auto} (a) the absolute value $|a(t)|$ of the
wavepacket autocorrelation function calculated following vertical
excitation to the $1B_{2u}(\pi\pi^{*})$ state of pyrazine with the
same four-state, seven-mode model used in the spectrum
simulations. Shown in Figure \ref{fig:pyrazine_auto} (b) are the
differences in $|a(t)|$, relative the the exact overlap results,
calculated using various levels of approximation.

\begin{figure}
  \begin{center}
    \includegraphics[width=7.0cm,angle=0]{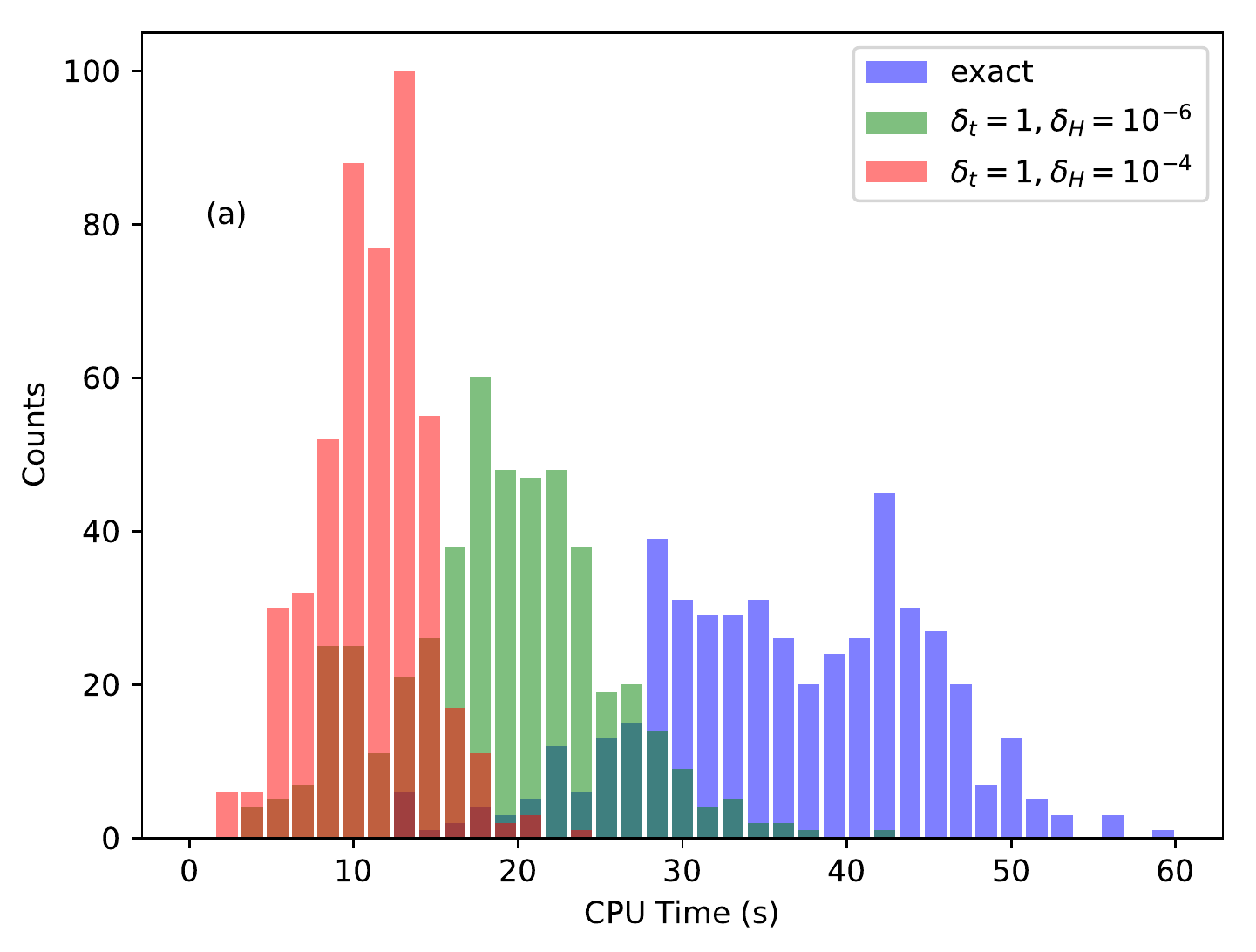}
    \includegraphics[width=7.0cm,angle=0]{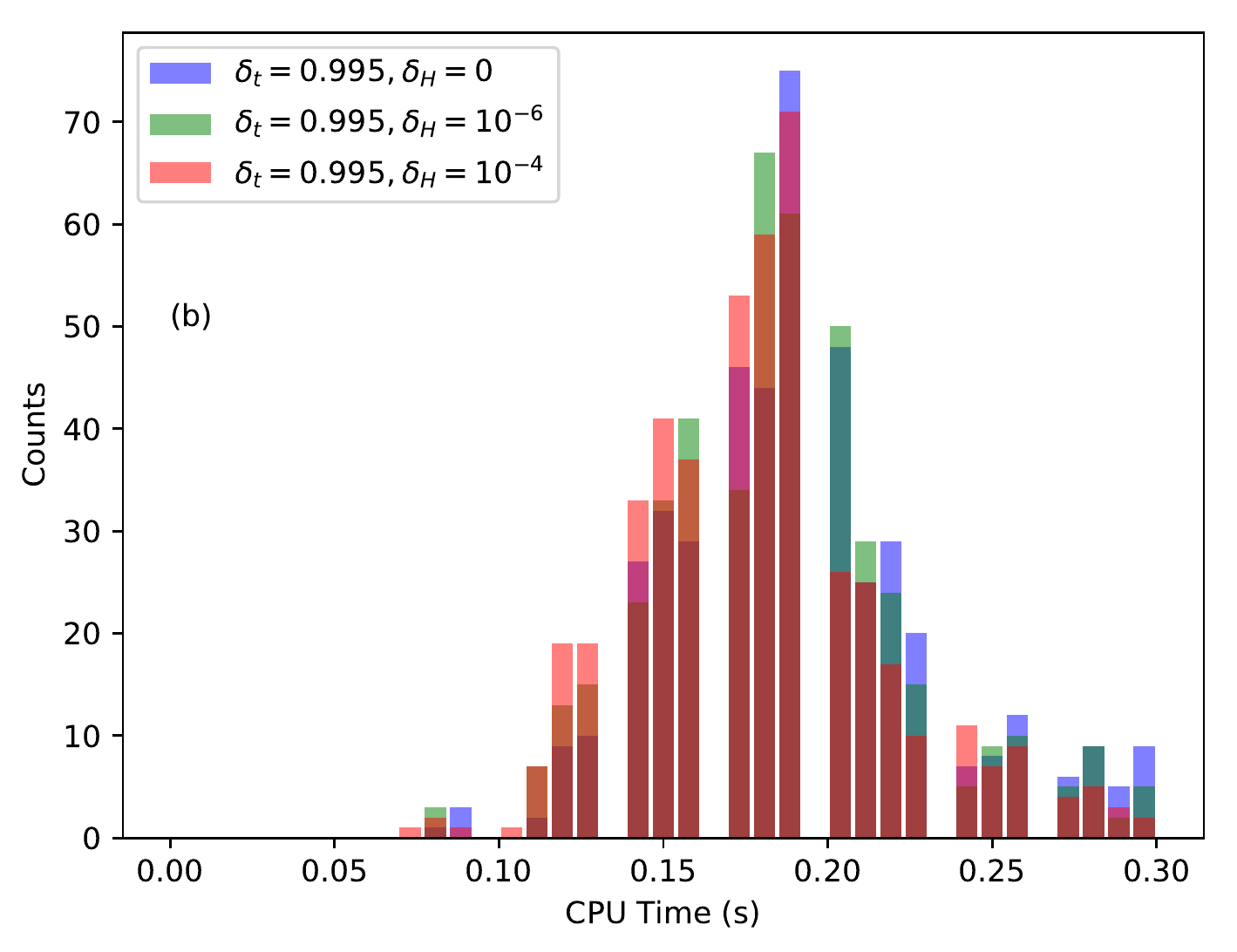}
    \caption{Timings for the P-BDD calculations for pyrazine using
      four electronic states calculated at the DFT/MRCI level of
      theory using the TZVP basis. (a) Timings for different screening
      thresholds $\delta_{H}$ and untruncated wavefunctions. (b)
      Timings for wavefunctions truncated using a norm truncation
      threshold $\delta_{t}=0.995$ and different screening thresholds
      $\delta_{H}$. All timings correspond to the use a single Intel
      i7-6700K CPU core.}
    \label{fig:bdd_timing}
  \end{center}
\end{figure}

We first consider the effects of Hadamard screening in the calculation
of wavefunction overlaps in the P-BDD calculations. Two levels of
approximations were used, corresponding to Hadamard screening
thresholds $\delta_{H}$ of $10^{-6}$ and $10^{-4}$. As can be seen in
Figure \ref{fig:pyrazine_auto} (b), the effect of using a Hadamard
screening threshold $\delta_{H}=10^{-6}$ has an effect on the
autocorrelation function that is of the order of machine
precision. Increasing $\delta_{H}$ to $10^{-4}$ increases the error in
the autocorrelation function, but the maximum error for times up to
240 fs is only $\mathcal{O}(10^{-6})$, which is still negligible. The
Hadamard screening of unique factors in the P-BDD calculations does,
however, result in savings in computational costs, as shown in Figure
\ref{fig:bdd_timing} (a). Here, the timings for the P-BDD calculations
are shown for the exact calculations (no screening), and Hadamard
screening thresholds of $\delta_{H}=10^{-6}$ and
$\delta_{H}=10^{-4}$. Note that because the DFT/MRCI method employs an
adaptive configuration selection algorithm\cite{grimme_dft-mrci}, the
numbers of Slater determinants, and hence the computational cost of
each P-BDD calculation, varies from geometry to geometry. It can
clearly be seen that the use of Hadamard screening leads to
significant savings in computational costs: the average timing for the
P-BDD calculations performed with no screening, $\delta_{H}=10^{-6}$
and $\delta_{H}=10^{-4}$ are 39, 19 and 11 seconds,
respectively. Considering the negligible loss in accuracy resulting
from the Hadamard screening procedure, these can be deemed to be
worthwhile savings.

Secondly, we examine the effects of wavefunction truncation in the
P-BDD calculations on the wavepacket autocorrelation function. To do
so, the P-BDD calculations were performed using wavefunction
expansions truncated to give wavefunction norms $||\psi_{I}||$ of
0.995. The effect of wavefunction truncation on the error in the
autocorrelation function is more pronounced than for the introduction
of Hadamard screening, as can be seen in Figure
\ref{fig:pyrazine_auto} (b). However, even up to 240 fs, the errors in
the autocorrelation function are still relatively small. Also shown
here are the errors in the autocorrelation function obtained when
using a combination of wavefunction truncation and Hadamard
screening. As may be expected, the errors resulting from the
combination of wavefunction truncation and Hadamard screening are of
the same order of magnitude as for wavefunction truncation
alone. Finally, we show in Figure \ref{fig:bdd_timing} (b) the timings
for the P-BDD calculations performed using truncated wavefunction
expansions. Relative to the untruncated results (shown in Figure
\ref{fig:bdd_timing} (a)), the results are striking, with a speed up
by two orders of magnitude being attained, and all calculations now
taking less than 0.3 seconds. In fact, with a norm threshold of 0.995,
the calculation of the wavefunction overlaps no longer dominates the
total P-BDD calculation, as can be seen by the very similar timings
for the $\delta_{H}=0$, $\delta_{H}=10^{-6}$, and $\delta_{H}=10^{-4}$
calculations.

\begin{figure}
  \begin{center}
    \includegraphics[width=7.0cm,angle=0]{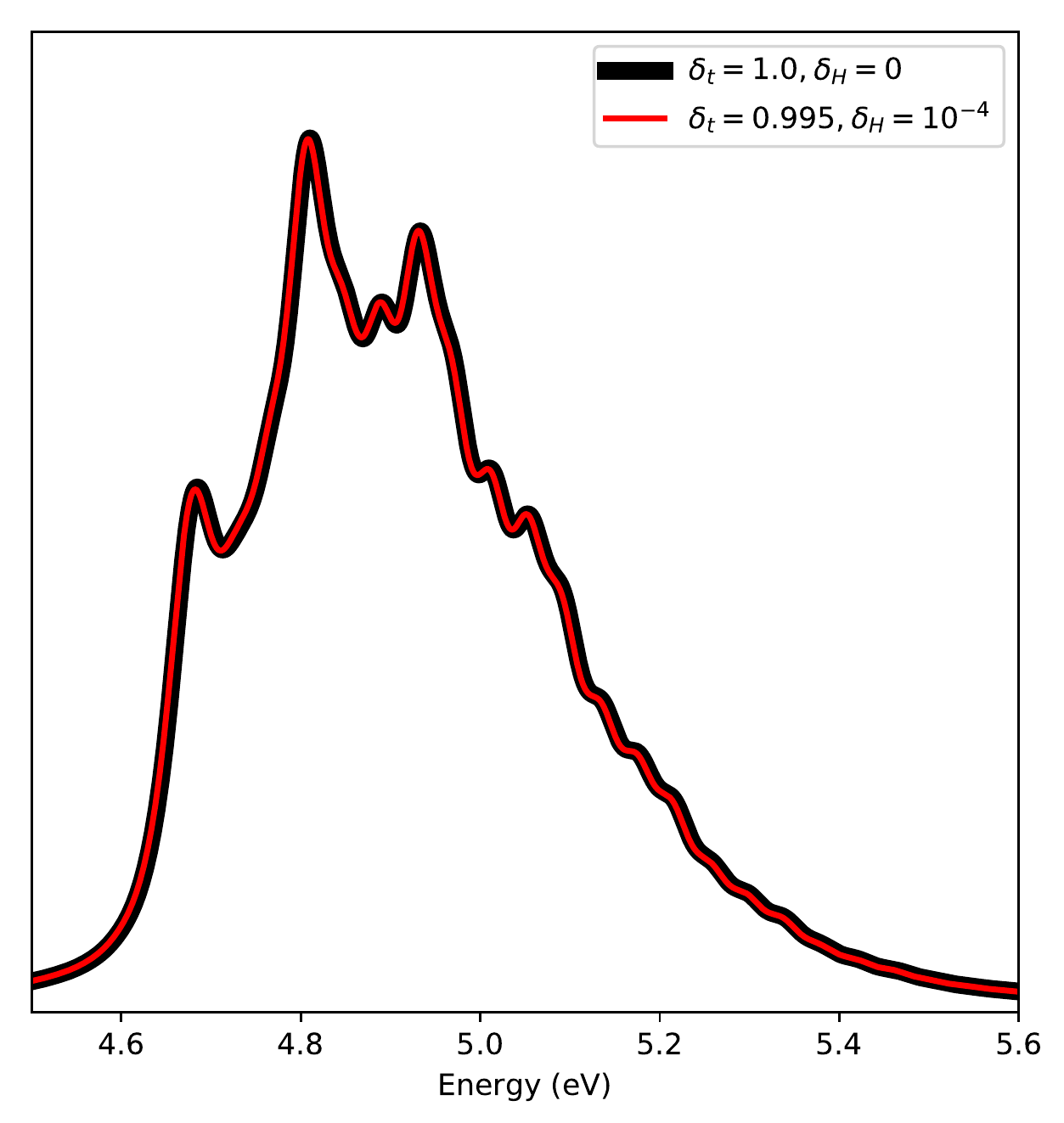}
    \caption{Absorption spectrum corresponding to vertical excitation
      to the $1B_{2u}(\pi\pi^{*})$ state of pyrazine calculated using
      the four-state, seven-mode model and different levels of
      approximation in the P-BDD diabatisation calculations. Thick
      black line: exact P-BDD calculations. Thin red line: norm
      truncation threshold $\delta_{t}=0.995$ and Hadamard screening
      threshold $\delta_{H}=10^{-4}$.}
    \label{fig:pyrazine_spec_approx}
  \end{center}
\end{figure}

Clearly, the use of Hadamard screening and wavefunction truncation
both lead to considerable speedups in a P-BDD calculation. In the case
of Hadamard screening, we observe that even a relatively loose
screening threshold ($\delta_{H}=10^{-4}$) leads to decreases in
computation cost of around $70\%$ with negligible effects on the
accuracy of the diabatic potentials. Wavefunction truncation leads to
even greater computational savings, by orders of magnitude. Here, the
degradation of the accuracy of the P-BDD diabatic potentials is found
to be greater, but is found to still be small enough to justify the
use of this approximation. To demonstrate this, we show in Figure
\ref{fig:pyrazine_spec_approx} the absorption spectrum calculated
using model potentials derived from P-BDD calculations employing no
approximations ($\delta_{t}=1.0$, $\delta_{H}=0$) and with both
norm truncation and a loose Hadamard screening threshold
($\delta_{t}=0.995$, $\delta_{H}=10^{-4}$). The two spectra are found
to be barely distinguishable from each other, lending justification to
this level of approximation.

\section{Conclusions}
The DFT/MRCI method is a well established tool for the calculation of
the excited states of large molecular systems. The excellent
cost-to-accuracy ratio of DFT/MRCI makes it an appealing choice for
the construction of excited state potentials and non-adiabatic
couplings for use in quantum dynamics simulations. The lack of
analytical derivative couplings within DFT/MRCI means that one must
adopt a diabatic representation in order to achieve this. The
challenge, then, is to determine a diabatisation procedure that is
compatible with DFT/MRCI. Due to the lack of analytical derivative
couplings and the constraint of having to use canonical KS MOs as the
single-particle basis, this a non-trivial task.

Our solution to this problem is to use a propagative variant of the
BDD method of Pacher, Cederbaum and
K\"{o}ppel\cite{pacher_bdd_1989,pacher_quasidiabatic,pacher_quasidiabatic_states_adv_chem_phys},
termed P-BDD. The P-BDD method is both formally rigorous and requires
as input only the overlaps of sets of electronic wavefunctions at
neighboring nuclear geometries, which are amenable to computation at
the DFT/MRCI level of theory.

The bottleneck in a P-BDD calculation is the determination of large
numbers of wavefunction overlaps in terms of non-orthogonal sets of
MOs. To render this computationally tractable, we implemented the
recently reported algorithm of Plasser {\it et
  al.}\cite{plasser_wf_overlaps} with two key modifications. Firstly,
common factors (unique determinants of spin orbital overlaps) were
determined and stored not just on a wavefunction-pair basis, but for
all wavefunction pairs simultaneously. Secondly, the use of Hadamard
screening of the unique factors was introduced. Using these
modifications, large numbers of wavefunction overlaps can be computed
efficiently, making the P-BDD procedure tractable even for large
molecules and many electronic states.

As an initial application of the DFT/MRCI P-BDD method, diabatic
potentials were calculated for LiH and pyrazine. For LiH, the DFT/MRCI
P-BDD diabatic potentials were found to correctly capture the strong
non-adiabatic coupling between the $1^{1}\Sigma^{+}$ and
$2^{1}\Sigma^{+}$ states. In particular, the derivative couplings
derived from the DFT/MRCI P-BDD diabatic potential matrix were found
to compare well to those calculated analytically using canonical MRCI
calculations. Additionally, a diabatic natural orbital analysis
revealed that the ionic and covalent characters of the $\tilde{X}$ and
$\tilde{A}$ diabatic states were maintained as the Li-H bond is
stretched, as should be the case for good diabatic states. For
pyrazine, a model vibronic coupling Hamiltonian was constructed by
direct fitting to DFT/MRCI P-BDD diabatic potentials. The model
potential was subsequently used in MCTDH quantum dynamics simulations
to compute the $1B_{2u}(\pi\pi^{*})$ state vibronic absorption
spectrum. Almost quantitative agreement between the simulated and
experimental spectra was attained, providing further validation of the
accuracy of the DFT/MRCI P-BDD diabatic potentials.

We close by noting that the ability to rapidly and accurately
determine diabatic potentials will be directly applicable to
on-the-fly quantum dynamics simulations. In particular, the
combination of DFT/MRCI P-BDD calculations and on-the-fly machine
learning\cite{richings_krr_standard_method,richings_krr_mctdh,richings_krr_mctdh_svd,richings_krr_mctdh_tensor_decomposition,polyak_gpr_ddvmcg}
holds promise for the construction of a powerful, near-black box
framework for the use of DFT/MRCI in on-the-fly dynamics calculations,
and will be the focus of future work in our group.

\section{Acknowledgments}
We would like to thank Christel Marian and Martin Kleinschmidt for
providing the original DFT/MRCI code that was modified for use in this
work.

\newpage
\clearpage
\appendix

\section{Normal equations fitting of the model potentials}\label{app:fitting}
The fitting of the one- and two-mode terms in Equation
\ref{eq:modpot}, $\tau_{p\alpha}^{(I,J)}$ and $\eta_{\alpha
  \beta}^{(I,J)}$, respectively, were fitted sequentially, starting
with the one-mode terms.

\subsection{Fitting of the one-mode terms}
Let $\boldsymbol{Q}_{0}^{\alpha}$ be the vector $\boldsymbol{Q}_{0}$
with the $\alpha$th element removed, and
$W_{IJ}(Q_{\alpha},\boldsymbol{Q}_{0}^{\alpha})$ denote the value of
the quasi-diabatic potential matrix element $W_{IJ}$ displaced only
along the single mode $Q_{\alpha}$ at the geometry
$(0,\dots,0,Q_{\alpha},0,\dots,0)$. Our model potential for
displacements along a single mode $Q_{\alpha}$ reads

\begin{equation}
  W_{IJ}^{mod}(Q_{\alpha},\boldsymbol{Q}_{0}^{\alpha}) = \tau_{0}^{(I,J)}
  + \sum_{p=1}^{4} \frac{1}{p!} \tau_{p\alpha}^{(I,J)} Q_{\alpha}^{p}.
\end{equation}

\noindent
Let $\{ W_{IJ}(Q_{\alpha},\boldsymbol{Q}_{0}^{\alpha})_{i} :
i=1,\dots,n \}$ denote the set of `true' quasi-diabatic potential
matrix element values calculated at a number, $n$, of different
geometries $(Q_{\alpha},\boldsymbol{Q}_{0}^{\alpha})_{i}$ with only
the mode $Q_{\alpha}$ displaced. The squared one-mode residuals are
defined as

\begin{equation}
  \begin{aligned}
    \left|R_{\alpha}^{(I,J)}\right|^{2} &\equiv \sum_{i=1}^{n} \Big[
      W_{IJ}(Q_{\alpha},\boldsymbol{Q}_{0}^{\alpha})_{i} -
      \tau_{0}^{(I,J)} \\
      &- \sum_{p=1}^{4} \frac{1}{p!}  \tau_{p\alpha}^{(I,J)}
      (Q_{\alpha}^{(i)})^{p} \Big]^{2},
  \end{aligned}
\end{equation}

\noindent
where $Q_{\alpha}^{(i)}$ is the value of $Q_{\alpha}$ in
$(Q_{\alpha},\boldsymbol{Q}_{0}^{\alpha})_{i}$. Requiring that the
partial derivatives of the squared residual $|R_{\alpha}^{(I,J)}|^{2}$
with respect to the parameters $\tau_{0}^{(I,J)}$ and
$\tau_{p\alpha}^{(I,J)}$ vanish leads to the following system of
linear equations:

\begin{equation}
  \boldsymbol{X}_{\alpha} \boldsymbol{t}_{\alpha}^{(I,J)} =
  \boldsymbol{w}_{\alpha}^{(I,J)},
\end{equation}

\noindent
with

\begin{equation}
  \boldsymbol{X}_{\alpha} =
  \left(
    \begin{array}{cccc}
      1 & Q_{\alpha}^{(1)} & \cdots & (Q_{\alpha}^{(1)})^{4} \\
      1 & Q_{\alpha}^{(2)} & \cdots & (Q_{\alpha}^{(2)})^{4} \\
      \vdots & \vdots & \ddots & \vdots \\
      1 & Q_{\alpha}^{(n)} & \cdots & (Q_{\alpha}^{(n)})^{4} \\
    \end{array}
    \right),
\end{equation}

\begin{equation}
  \boldsymbol{t}_{\alpha}^{(I,J)} =
  \left(
    \begin{array}{c}
      \tau_{0}^{(I,J)} \\
      \frac{1}{1!} \tau_{1\alpha}^{(I,J)} \\
      \vdots \\
      \frac{1}{4!} \tau_{4\alpha}^{(I,J)} \\
    \end{array}
    \right),
\end{equation}

\begin{equation}
  \boldsymbol{w}_{\alpha}^{(I,J)} = 
  \left(
    \begin{array}{c}
      W_{IJ}(Q_{\alpha},\boldsymbol{Q}_{0}^{\alpha})_{1} \\
      W_{IJ}(Q_{\alpha},\boldsymbol{Q}_{0}^{\alpha})_{2} \\
      \vdots \\
      W_{IJ}(Q_{\alpha},\boldsymbol{Q}_{0}^{\alpha})_{n} \\
    \end{array}
    \right)
\end{equation}

\noindent
This yields the following solution for the optimal set of one-mode
coefficients:

\begin{equation}
  \boldsymbol{t}_{\alpha}^{(I,J)} = \left( \boldsymbol{X}_{\alpha}^{T}
  \boldsymbol{X}_{\alpha} \right)^{-1} \boldsymbol{X}_{\alpha}^{T}
  \boldsymbol{w}_{\alpha}^{(I,J)}.
\end{equation}

In the fitting of the one-mode terms, for each mode $Q_{\alpha}$
quasi-diabatic potential matrices $\boldsymbol{W}(\boldsymbol{Q})$
were calculated at 21 geometries
$(Q_{\alpha},\boldsymbol{Q}_{0}^{\alpha})_{i}=(i\Delta
Q_{\alpha},\boldsymbol{Q}_{0}^{\alpha})$, $\Delta Q_{\alpha}=0.5$,
$i=-10,\dots,0,\dots,10$.

\subsection{Fitting of the two-mode terms}
Once the one-mode terms $\tau_{p\alpha}^{(I,J)}$ have been determined,
the two-mode terms $\eta_{\alpha \beta}^{(I,J)}$ can be calculated from
fitting to the values of the diabatic potential matrix elements
$W_{IJ}(\boldsymbol{Q})$ calculated at geometries
$\boldsymbol{Q}=(Q_{\alpha},Q_{\beta},\boldsymbol{Q}_{0}^{\alpha\beta})$
with two, and only two, modes displaced. Here,
$\boldsymbol{Q}_{0}^{\alpha\beta}$ denotes the vector
$\boldsymbol{Q}_{0}$ with the $\alpha$th and $\beta$th elements
removed, and $(Q_{\alpha},Q_{\beta},\boldsymbol{Q}_{0}^{\alpha\beta})$
is to be equated with the vector
$(0,\dots,0,Q_{\alpha},0,\dots,0,Q_{\beta},0\dots,0)$. The part of
$W_{IJ}(\boldsymbol{Q})$ corresponding to the two-body correlation of
the modes $Q_{\alpha}$ and $Q_{\beta}$, denoted by
$\tilde{W}_{IJ}^{\alpha\beta}(Q_{\alpha},Q_{\beta})$, can be obtained as

\begin{equation}
  \begin{aligned}
    \tilde{W}_{IJ}^{\alpha\beta}(Q_{\alpha},Q_{\beta}) &=
    W_{IJ}(Q_{\alpha},Q_{\beta},\boldsymbol{Q}_{0}^{\alpha\beta}) \\
    &- W_{IJ}(\boldsymbol{Q}_{0}) -
    W_{IJ}(Q_{\alpha},\boldsymbol{Q}_{0}^{\alpha}) -
    W_{IJ}(Q_{\beta},\boldsymbol{Q}_{0}^{\beta}).
  \end{aligned}
\end{equation}

\noindent
In our model potential,
$\tilde{W}_{IJ}^{\alpha\beta}(Q_{\alpha},Q_{\beta})$ is approximated
as

\begin{equation}
  \tilde{W}_{IJ}^{\alpha\beta}(Q_{\alpha},Q_{\beta}) \approx
  \eta_{\alpha\beta}^{(I,J)} Q_{\alpha} Q_{\beta}.
\end{equation}

\noindent
Thus, we minimise the following squared two-mode residuals with
respect to the two-mode terms $\eta_{\alpha\beta}^{(I,J)}$:

\begin{equation}
  \left|R_{\alpha\beta}^{(I,J)}\right|^{2} \equiv \sum_{i=1}^{n}
  \left[ \tilde{W}_{IJ}^{\alpha\beta}(Q_{\alpha},Q_{\beta})_{i} -
    \eta_{\alpha\beta}^{(I,J)} Q_{\alpha}^{(i)} Q_{\beta}^{(i)}
    \right]^{2},
\end{equation}

\noindent
where $Q_{\alpha}^{(i)}$ and $Q_{\beta}^{(i)}$ are the values of
$Q_{\alpha}$ and $Q_{\beta}$ in the set of $n$ geometries $\{
(Q_{\alpha},Q_{\beta},\boldsymbol{Q}_{0}^{\alpha\beta})_{i} \}$ at
which the values
$\tilde{W}_{IJ}^{\alpha\beta}(Q_{\alpha},Q_{\beta})_{i}$ are
calculated.

Minimisation of the squared residual $|R_{\alpha\beta}^{(I,J)}|^{2}$
with respect to the parameter $\eta_{\alpha\beta}^{(I,J)}$ yields the
following equation:

\begin{equation}\label{eq:2-mode_terms}
  \eta_{\alpha\beta}^{(I,J)} = \frac{\sum_{i=1}^{n}
    \tilde{W}_{IJ}^{\alpha\beta}(Q_{\alpha},Q_{\beta})_{i}}{\sum_{i=1}^{n}
    Q_{\alpha}^{(i)} Q_{\beta}^{(i)}}
\end{equation}

\noindent
Care must be taken here in the selection of the geometries
$(Q_{\alpha},Q_{\beta},\boldsymbol{Q}_{0}^{\alpha\beta})_{i}$ to be
included in the fitting process as an improper selection will lead to
the denominator in Equation \ref{eq:2-mode_terms} vanishing. Our
choice corresponds to the diagonal cuts
$(Q_{\alpha},Q_{\beta},\boldsymbol{Q}_{0}^{\alpha\beta})_{i} = (i
\Delta Q, i \Delta Q, \boldsymbol{Q}_{0}^{\alpha\beta})$, $\Delta
Q=0.5$, $i=-2,-1,0,1,2$.

Finally, we note that in our fitting scheme we approximate
$\tilde{W}_{IJ}^{\alpha\beta}(Q_{\alpha},Q_{\beta})$ using the prior
fitted one-mode terms $\tau_{p\alpha}^{(I,J)}$:

\begin{equation}
  \begin{aligned}
    \tilde{W}_{IJ}^{\alpha\beta}(Q_{\alpha},Q_{\beta}) &\approx 
    W_{IJ}(Q_{\alpha},Q_{\beta},\boldsymbol{Q}_{0}^{\alpha\beta}) \\
    &- \tau_{0}^{(I,J)} - \sum_{p=1}^{4} \left( \tau_{p\alpha}^{(I,J)}
    Q_{\alpha}^{p} + \tau_{p\beta}^{(I,J)} Q_{\beta}^{p} \right).
  \end{aligned}
\end{equation}

\noindent
For small displacements this will be a valid approximation.

\subsection{Symmetry}
It remains to note that many of the expansion coefficients
$\tau_{p\alpha}^{(I,J)}$ and $\eta_{\alpha \beta}^{(I,J)}$ will be zero by
symmetry. Specifically, the following relations hold:

\begin{equation}
  \tau_{p\alpha}^{(I,J)} \ne 0, \hspace{0.5cm} \left\{
  \bigotimes_{K=1}^{p} \Gamma^{\alpha} \right\} \otimes \Gamma^{I}
  \otimes \Gamma^{J} \ni \Gamma^{1}.
\end{equation}

\begin{equation}
  \eta_{\alpha\beta}^{(I,J)} \ne 0, \hspace{0.5cm} \Gamma^{\alpha}
  \otimes \Gamma^{\beta} \otimes \Gamma^{I} \otimes \Gamma^{J} \ni
  \Gamma^{1},
\end{equation}

\noindent
where $\Gamma^{\alpha}$ and $\Gamma^{I}$ denote the irreducible
representations generated by the mode $Q_{\alpha}$ and the state
$\phi_{I}$, respectively, and $\Gamma^{1}$ the totally symmetric
irreducible representation of the point group in question.

While it would be possible to omit from the fitting procedure those
expansion coefficients that are zero by symmetry, we instead choose to
include them and to monitor the fitted values of these. This allows
for one to spot any possible symmetry breaking in the P-BDD
calculations. A maximum value of $3 \times 10^{-5}$ eV was attained
for a coefficient that was zero by symmetry. All such parameters were
omitted in the resulting MCTDH calculations, but this analysis does
serve to demonstrate the symmetry-preserving properties of the P-BDD
procedure.


\end{document}